\documentclass[a4paper,11pt]{article}
\pdfoutput=1 

\usepackage{jcappub} 

\usepackage[T1]{fontenc} 

\usepackage{amsmath}
\usepackage{amssymb}
\usepackage{color}
\usepackage[varg]{txfonts}
\usepackage{bm}

\newcommand{\calR}{{\cal R}}
\title{\boldmath Non-Gaussianity effects on the primordial black hole abundance for sharply-peaked primordial spectrum}

\author[a,b]{Takahiko Matsubara}
\author[c,d,e]{and Misao Sasaki}

\affiliation[a]{Institute of Particle and Nuclear Studies, High Energy
  Accelerator Research Organization (KEK), Oho 1-1, Tsukuba 305-0801,
  Japan}

\affiliation[b]{The Graduate University for Advanced Studies
  (SOKENDAI), Tsukuba 305-0801, Japan}

\affiliation[c]{Kavli Institute for the Physics and Mathematics of the
  Universe (WPI), University of Tokyo, Chiba 277-8583, Japan}

\affiliation[d]{Center for Gravitational Physics and Quantum
  Information, Yukawa Institute for Theoretical Physics, Kyoto
  University, Kyoto 606-8502, Japan}

\affiliation[e]{Leung Center for Cosmology and Particle Astrophysics,
  National Taiwan University, Taipei 10617, Taiwan}

\emailAdd{tmats@post.kek.jp}
\emailAdd{misao.sasaki@ipmu.jp}

\abstract{ We perturbatively study the effect of non-Gaussianities on
  the mass fraction of primordial black holes (PBHs) at the time of
  formation by systematically taking its effect into account in the
  one-point probability distribution function of the primordial
  curvature perturbation. We focus on the bispectrum and trispectrum
  and derive formulas that describe their effects on the skewness and
  kurtosis of the distribution function. Then considering the case of
  narrowly peaked spectra, we obtain simple formulas that concisely
  express the effect of the bi- and trispectra. In particular,
  together with the $g_{\rm NL}$ and $\tau_{\rm NL}$ parameters of the
  trispectrum, we find that non-Gaussianity parameters for various
  types of the bispectrum are linearly combined to give an effective
  parameter, $f_{\rm NL}^{\rm eff}$, that determines the PBH mass
  fraction in the narrow spectral shape limit. }

\begin{document}
\maketitle
\flushbottom

\section{Introduction}

The possibility that black holes may be formed in the very early
universe was suggested about half a century ago
\cite{Hawking:1971ei,Carr:1974nx,Carr:1975qj}. Since then it has been
a topic of constant interest in cosmology, but it has never been
explored in depth. However, thanks to the rapidly growing interest in
gravitational wave cosmology in recent years, as well as to the
theoretical progress and technical developments, the primordial black
holes (PBHs) have become one of the hottest topics in cosmology today
(for recent reviews, see e.g., \cite{Carr:2020xqk,Carr:2021bzv}).

One of the most studied PBH formation mechanisms is the collapse of a
region with a sufficiently large curvature perturbation, presumably
produced from inflation, during the radiation-dominated early
universe. While such high peaks in the curvature perturbation hardly
exist in the conventional models of inflation, since there are
virtually no stringent observational constraint on cosmologically very
small scales that are relevant for the PBH formation, various models
that can produce sufficiently abundant PBHs have been discussed in the
literature.

The abundance of PBHs produced in the early Universe plays a crucial
role in the present Universe. Depending on the typical mass of PBHs,
various effects which are observationally detectable are anticipated
if the abundance is sufficiently large. There is no compelling
evidence for the existence of PBHs at any mass scale up to the present
time. Hence, what we have so far is the upper limit in the PBH
abundance in broad ranges of mass scale (For a recent review, see
Ref.~\cite{Carr:2020gox}). If PBHs are formed from high peaks of the
curvature perturbation, an upper bound on the PBH abundance places an
upper bound on the amplitude of the primordial power spectrum. The
observed amplitude of the power spectrum on scales of the cosmic
microwave background (CMB) and the large-scale structure (LSS) is
too small to produce PBHs. However, scales relevant to the PBH
formation are typically much smaller than those scales. Thus, various
mechanisms that give rise to the amplitude of the power spectrum on
small scales much larger than that on the CMB scale have been proposed
(see, e.g., Ref.~\cite{Green:2014faa}). In many of those models,
however, the non-Gaussianities in the curvature perturbation are
negligible or simply ignored. Nevertheless, even if they are small,
they may considerably affect the PBH formation, as it is acutely
sensitive to non-Gaussian features in the tail of the probability
distribution function \cite{Byrnes:2012yx}.

The effects of non-Gaussianity upon the abundance of PBHs have been
investigated by many authors using various methods
\cite{Bullock:1996at,PinaAvelino:2005rm,Seery:2006wk,Shandera:2012ke,Young:2013oia,Young:2015cyn,Atal:2019cdz,Yoo:2019pma,Kitajima:2021fpq,Taoso:2021uvl,Riccardi:2021rlf,Pi:2021dft,Young:2022phe,Kehagias:2019eil}.
However, in most of the previous work, only the local-type of
non-Gaussianity is considered. Even for the bispectrum, which is the
lowest non-trivial order in perturbation where the non-Gaussianity
appears, only a limited number of authors consider the other types,
such as equilateral- and orthogonal-types of
non-Gaussianity~\cite{Shandera:2012ke,Young:2015cyn}.
The effect of the trispectrum of non-Gaussianity on the abundance have
not been discussed much, while the clustering of PBHs in the presence
of local-type trispectra are considered in, e.g.,
Refs.~\cite{Tada:2015noa,Suyama:2019cst,Matsubara:2019qzv}. Thus, it
is desirable to understand the effect of various types of
non-Gaussianity on the PBH formation more systematically.

In this paper, we analytically study the effect of non-Gaussianity up
through the trispectrum in a model-independent way as much as
possible. To analytically address the problem, we use a set
of approximations that are valid in many situations. In the presence
of skewness and kurtosis in the primordial fluctuations with
hierarchical orders, we derive a formula for the non-Gaussian
corrections to the abundance of PBHs in a high-peaks limit,
generalizing the known formula for the threshold. We also derive
integral formulas to calculate the values of skewness and kurtosis in
popular models of non-Gaussianity, including local-, equilateral-,
folded- and orthogonal-type models for bispectrum, and generalized
local-type models for trispectrum. These integral formulas are further
reduced to asymptotic formulas of analytic forms by taking the limit
of a sharply-peaked shape of the primordial spectrum. The derived
formulas provide useful apparatus for predicting the abundance of PBHs
in a variety of models.

This paper is organized as follows: In section~\ref{sec:Abundance},
after the basic notations and definitions we use in this paper are given,
high-peaks formulas for the PBH formation are introduced. In
section~\ref{sec:Cumulants}, we derive integral formulas for skewness
and kurtosis parameters for a class of non-Gaussian models mentioned
above, for an arbitrary shape of the primordial spectrum. In
section~\ref{sec:Narrow}, the derived integral formulas are reduced to
analytic forms by taking the sharply-peaked limit of the spectrum. In
section~\ref{sec:Numrical}, we numerically evaluate the behavior of
derived formulas and demonstrate it in several cases.

\section{\label{sec:Abundance}
  The abundance of PBHs from non-Gaussian initial conditions}

\subsection{Definitions}

First, we define the fundamental quantities used in this paper. The PBHs
are considered to be formed from large positive perturbations of the
3-dimensional curvature $\delta R^{(3)}$ at an early stage of the
Universe, well before the astrophysical structure formation takes
place. The 3-dimensional curvature perturbation on comoving slices is
characterized by the curvature perturbation in Fourier space, denoted
by $\calR(\bm{k})$. In linear order, we have
$\delta R^{(3)}=4(k^2/a^2)\calR$, where $a$ is the scale factor. Since
the Einstein equations imply $\delta R^{(3)}\approx 6H^2\varDelta$ on
or above Hubble scales at linear order, where $H=\dot{a}/a$ is the
Hubble parameter and $\varDelta$ is the energy density contrast on
comoving slices, large positive 3-dimensional curvature perturbations
are equivalent to large positive energy density fluctuations.

If its probability distribution is Gaussian, the statistical
properties of the comoving curvature perturbation are completely
characterized by the power spectrum $P_\calR(k)$. The power spectrum
is the two-point correlation of the fluctuations in Fourier space,
and is defined by
\begin{equation}
  \langle\calR(\bm{k}_1)\calR(\bm{k}_2)\rangle_\mathrm{c}
  = (2\pi)^3 \delta_\mathrm{D}^3(\bm{k}_1 + \bm{k}_2) P_\calR(k_1),
  \label{eq:2-01a}
\end{equation}
where $\delta_\mathrm{D}^3(\bm{k})$ is the Dirac's delta function and
$\langle\cdots\rangle_\mathrm{c}$ denotes the cumulants or connected part of the statistical average. For the two-point cumulant above, it can be replaced by a simple average, assuming the mean value of curvature fluctuations is zero, $\langle\calR\rangle
= 0$.

The presence of non-Gaussianity gives rise
to higher-order correlations, such as the bispectrum, trispectrum, and
so forth. The bispectrum $B_\calR(\bm{k}_1,\bm{k}_2,\bm{k}_3)$ and the
trispectrum $T_\calR(\bm{k}_1,\bm{k}_2,\bm{k}_3,\bm{k}_4)$ are three- and four-point functions in Fourier space, defined by
\begin{align}
    \langle\calR(\bm{k}_1)\calR(\bm{k}_2)\calR(\bm{k}_3)\rangle_\mathrm{c}
    &= (2\pi)^3 \delta_\mathrm{D}^3(\bm{k}_1 + \bm{k}_2 + \bm{k}_3)
    B_\calR(\bm{k}_1,\bm{k}_2,\bm{k}_3),
  \label{eq:2-01b}\\
  \langle\calR(\bm{k}_1)\calR(\bm{k}_2)\calR(\bm{k}_3)
  \calR(\bm{k}_4)\rangle_\mathrm{c}
    &= (2\pi)^3
      \delta_\mathrm{D}^3(\bm{k}_1 + \bm{k}_2 + \bm{k}_3 + \bm{k}_4)
      T_\calR(\bm{k}_1,\bm{k}_2,\bm{k}_3,\bm{k}_4).
  \label{eq:2-01c}
\end{align}
The three-point cumulant in Eq.~(\ref{eq:2-01b}) again can be replaced
by simple average because the mean value is zero. However, the
cumulant in Eq.~(\ref{eq:2-01c}) cannot be replaced by simple average,
and we have
$\langle\calR_1\calR_2\calR_3\calR_4\rangle_\mathrm{c} =
\langle\calR_1\calR_2\calR_3\calR_4\rangle -
\langle\calR_1\calR_2\rangle\langle \calR_3\calR_4\rangle -
\langle\calR_1\calR_3\rangle\langle \calR_2\calR_4\rangle -
\langle\calR_1\calR_4\rangle\langle \calR_2\calR_3\rangle$. 
The appearance of Dirac's delta functions in the above definition is due to the assumed statistical homogeneity 
in 3-space with translational symmetry. 

In general, in linear order, the relation between the comoving
curvature perturbation $\calR$ and the density contrast $\varDelta$
on comoving slices is given by
\cite{Liddle:2000cg,Green:2004wb,Young:2014ana}
\begin{equation}
  \label{eq:2-02}
  \varDelta(\bm{k};t) = \mathcal{M}(k) \calR(\bm{k})\,,
\end{equation}
where
\begin{equation}
  \label{eq:2-03}
  \mathcal{M}(k;t) = \frac{2+2w}{5+3w}
  \left(\frac{k}{aH}\right)^2,
\end{equation}
and $w=p/\rho$ is the equation of state parameter. The PBH formation
criteria are typically described by smoothed density fields with a
smoothing radius of the horizon scale, $R=(aH)^{-1}$. In the
following, we assume the formation of PBHs takes place in a
radiation-dominated epoch, $w=1/3$, and thus we use
\begin{equation}
  \mathcal{M}(k) = \frac{4}{9} k^2R^2\,,
  \label{eq:2-04}
\end{equation}
for the coefficient of Eq.~(\ref{eq:2-02}).

The smoothed density field in configuration space is given by
\begin{equation}
  \label{eq:2-05}
  \varDelta_R(\bm{x}) = \int \frac{d^3k}{(2\pi)^3}
  e^{i\bm{k}\cdot\bm{x}} \varDelta(\bm{k}) W(kR)
\end{equation}
where $W(kR)$ is the window function for the smoothing. 
In this paper, we adopt the Gaussian window function,
\begin{equation}
  \label{eq:2-06}
  W(kR) = \exp\left(-\frac{k^2R^2}{2}\right)\,.
\end{equation}

\subsection{The abundance of PBHs} 

The abundance of PBHs is frequently modeled by the initial mass
fraction $\beta$ of the universe that turns into PBHs at the time of
formation. One of the simplest and most commonly used criteria for the
PBH formation is to set a threshold density contrast
$\varDelta_\mathrm{c}$. Then $\beta$ may be estimated by computing the
fraction of space with $\varDelta_R(\bm{x}) > \varDelta_\mathrm{c}$ in
the initial density field. This gives
\begin{equation}
  \beta_\mathrm{th} = \int_{\varDelta_\mathrm{c}}^\infty
  P(\varDelta_R;R)\,d\varDelta_R,
  \label{eq:3-01}
\end{equation}
where $P(\varDelta_R;R)$ is the probability distribution function of
the initial density field smoothed over a horizon scale $R$. In
the analogy to the Press-Schechter theory of structure formation
\cite{Press:1973iz}, a fudge factor 2 is sometimes put in front of the
right-hand side (RHS) of the above equation. Our discussion below does
not depend on whether this fudge factor is present or not. Another
criteria of the PBH formation is to use the peak theory
\cite{Bardeen:1985tr,Green:2004wb}. Given the number density of peaks
$n_\mathrm{pk}(\varDelta_\mathrm{c};R)$ of the smoothed density field
above the threshold, $\varDelta_\mathrm{c}$, the mass fraction is given
by
\begin{equation}
  \beta_\mathrm{pk} =
  (2\pi)^{3/2}R^3 n_\mathrm{pk}(\varDelta_\mathrm{c};R),
  \label{eq:3-02}
\end{equation}
where the prefactor $(2\pi)^{3/2} R^3$ corresponds to the effective
volume of the Gaussian filter.
There is yet another formation for the threshold based on the
so-called Compaction function
\cite{Musco:2020jjb,Musco:2018rwt,Kehagias:2019eil}. But as its relation to
the probability distribution function seems rather non-trivial, we
leave this case for future studies.

When the initial density field is given by a random Gaussian field,
the one-point probability distribution function is given by
$P = (2\pi {\sigma}^2)^{-1/2}e^{-{\varDelta_R}^2/(2{\sigma}^2)}$,
where ${\sigma}^2 = \langle{\varDelta_R}^2\rangle$ is the variance of
the smoothed density field. In this case, Eq.~(\ref{eq:3-01}) is given
by
\begin{equation}
  \beta_\mathrm{th}^\mathrm{G}
  = \frac{1}{2}
  \mathrm{erfc}\left(\frac{\nu}{\sqrt{2}}\right)
  \approx
  \frac{1}{\sqrt{2\pi}} \frac{e^{-\nu^2/2}}{\nu},
  \label{eq:3-03}
\end{equation}
where $\nu \equiv \varDelta_\mathrm{c}/\sigma$ is the normalized
threshold. The last expression of the above equation is an asymptotic
form for a high threshold of $\nu\gg 1$. The number density of peaks
in a random Gaussian field is calculated from the joint probability
distribution of field derivatives and is analytically given by an
integral form \cite{Bardeen:1985tr}. An asymptotic form of the result
for high peaks ($\nu\gg 1$) is given by
\begin{equation}
  n_\mathrm{pk} = \frac{1}{(2\pi)^2}
  \left(\frac{{\sigma_1}^2}{3{\sigma}^2}\right)^{3/2}
    (\nu^2-1) e^{-\nu^2/2},
  \label{eq:3-04}
\end{equation}
where
\begin{equation}
  {\sigma_j}^2 = \int \frac{k^2dk}{2\pi^2} k^{2j}P_\varDelta(k) W^2(kR)
  \label{eq:3-05}
\end{equation}
and $P_\varDelta(k)$ is the power spectrum of the density field
$\varDelta$. The corresponding expression for the mass fraction is
given by
\begin{equation}
  \beta_\mathrm{pk}^\mathrm{G} =
  \frac{1}{\sqrt{2\pi}}
  \left(\frac{R\sigma_1}{\sqrt{3}\sigma}\right)^3
    (\nu^2-1) e^{-\nu^2/2}.
  \label{eq:3-06}
\end{equation}

There are differences in the predictions of the two different models,
Eqs.~(\ref{eq:3-03}) and (\ref{eq:3-06}) with the Gaussian initial
conditions. However, the overall shape is relatively close to each
other if we take a higher threshold value $\nu$ for the peak theory
and the threshold theory for reasonable ranges in the PBH mass
\cite{Green:2004wb,Young:2014ana}.

\subsection{Non-Gaussian distributions}

The one-point distribution function of non-Gaussian fields are
characterized by higher-order cumulants
$\langle {\varDelta_R}^n \rangle_\mathrm{c}$. We assume a hierarchical
scaling of higher-order cumulants,
$\langle {\varDelta_R}^n \rangle_\mathrm{c} \propto {\sigma}^{2n-2}$,
and define reduced cumulants
\begin{equation}
  S_n \equiv 
  \frac{\langle {\varDelta_R}^n \rangle_\mathrm{c}}
  {{\sigma}^{2n-2}},
\label{eq:4-01}
\end{equation}
which are considered to be of order unity or less.
In general, the probability
distribution function in the integrand of Eq.~(\ref{eq:3-01}) can be
expanded in an Edgeworth series,
\begin{multline}
  P(\varDelta_R;R) =
  \frac{e^{-\nu^2/2}}{\sqrt{2\pi}\sigma}
  \Biggl\{ 1 +
    \frac{S_3}{3!} H_3(\nu) \sigma +
    \left[\frac{S_4}{4!} H_4(\nu) +
      \frac{1}{2!}\left(\frac{S_3}{3!}\right)^2 H_6(\nu)
    \right] \sigma^2
    \\
    + 
    \left[\frac{S_5}{5!} H_4(\nu) +
      \frac{S_3S_4}{3!4!} H_7(\nu) +
      \frac{1}{3!}\left(\frac{S_3}{3!}\right)^3 H_9(\nu)
    \right] \sigma^3 + \cdots
  \Biggr\},
  \label{eq:4-02}
\end{multline}
where $H_n(\nu) = e^{\nu^2/2}(-d/d\nu)^ne^{-\nu^2/2}$ is the Hermite
polynomial. In practice, one can truncate the series in some order
when the inequality $\sigma \ll \nu^{-3}$ is satisfied. The Edgeworth
expansion has been used to investigate the abundance of PBHs
\cite{Shandera:2012ke}. For high peaks, $\nu \gg 1$, the truncated
Edgeworth expansion is only applicable for sufficiently small
$\sigma$. Integrating the Edgeworth series above, the mass fraction in
the threshold model, Eq.~(\ref{eq:3-01}) reduces to
\begin{multline}
  \beta_\mathrm{th}
  = \beta_\mathrm{th}^\mathrm{G}
  + \frac{e^{-\nu^2/2}}{\sqrt{2\pi}}
  \Biggl\{
    \frac{S_3}{3!} H_2(\nu) \sigma +
    \left[\frac{S_4}{4!} H_3(\nu) +
      \frac{1}{2!}\left(\frac{S_3}{3!}\right)^2 H_5(\nu)
    \right] \sigma^2
    \\
    + 
    \left[\frac{S_5}{5!} H_3(\nu) +
      \frac{S_3S_4}{3!4!} H_6(\nu) +
      \frac{1}{3!}\left(\frac{S_3}{3!}\right)^3 H_8(\nu)
    \right] \sigma^3 + \cdots
  \Biggr\}.
  \label{eq:4-03}
\end{multline}

In the high-peaks limit, the mass fraction in the threshold model has
been derived in the context of biased structure formation, and the
result is given by \cite{Matarrese:1986et,Franciolini:2018vbk}
\begin{equation}
  \beta_\mathrm{th} \approx
  \frac{1}{\sqrt{2\pi}}
  \frac{e^{-\nu^2/2}}{\nu}
  \exp\left(
    \sum_{n=3}^\infty
    \frac{\nu^n}{n!}
    \frac{\langle {\varDelta_R}^n\rangle}{\sigma^n}
    \right) =
  \frac{\nu^{-1}}{\sqrt{2\pi}}
  \exp\left[
    -\frac{\nu^2}{2}
    \left(1-
      2\sum_{n=3}^\infty \frac{{\varDelta_\mathrm{c}}^{n-2}}{n!}S_n\right)
    \right].
  \label{eq:4-04}
\end{equation}
This result can also be obtained by summing up all the infinite series
of the leading contributions in the high-peaks limit
$H_n(\nu) \rightarrow \nu^n$ in the Eq.~(\ref{eq:4-03}). In this way,
the abundance of PBHs for non-Gaussian initial conditions in the
high-peaks limit is characterized by the series of reduced cumulants
$S_n$.

The above formula for the non-Gaussianity in the high-peaks limit can
be generalized to the peaks model of Eq.~(\ref{eq:3-06}). The details of the derivation are given in Appendix~\ref{app:A01}. 
One finds that the non-Gaussian contributions are the same as those in the high-peaks limit of the threshold model.
Namely,
\begin{equation}
  \beta_\mathrm{pk} \approx
  \frac{1}{\sqrt{2\pi}}
  \left(\frac{R\sigma_1}{\sqrt{3}\sigma}\right)^3
  (\nu^2-1)
  \exp\left[
    -\frac{\nu^2}{2}
    \left(1-
      2\sum_{n=3}^\infty \frac{{\varDelta_\mathrm{c}}^{n-2}}{n!}S_n\right)
    \right].
  \label{eq:4-05}
\end{equation}
Therefore, irrespective of the formation models of PBHs, the effect of non-Gaussianity in the high-peaks limit may be expressed in the generic form,
\begin{equation}
  \beta \approx
  \beta^\mathrm{G}
  \exp\left(\nu^2
    \sum_{n=3}^\infty \frac{{\varDelta_\mathrm{c}}^{n-2}}{n!}S_n
  \right)
  = A(\nu)
  \exp\left[
    -\frac{\nu^2}{2}
    \left(1-
      2\sum_{n=3}^\infty \frac{{\varDelta_\mathrm{c}}^{n-2}}{n!}S_n\right)
    \right],
  \label{eq:4-06}
\end{equation}
where
\begin{equation}
  \beta^\mathrm{G} = A(\nu)  e^{-\nu^2/2}
  \label{eq:4-07}
\end{equation}
is the mass fraction for the Gaussian initial condition, and $A(\nu)$ is the prefactor that depends on the formation models of PBHs, i.e., 
\begin{equation}
  A(\nu) \equiv
  \frac{1}{\sqrt{2\pi}} \times
  \begin{cases}
    \displaystyle
    \nu^{-1} & \mathrm{(threshold \ model)}, \\
    \displaystyle
    \left(\frac{R\sigma_1}{\sqrt{3}\sigma}\right)^3(\nu^2-1)
    & \mathrm{(peaks\ model)}.
  \end{cases}
  \label{eq:4-08}
\end{equation}
We note that the asymptotic formula~(\ref{eq:4-06}) is consistent only when
\begin{equation}
 2\sum_{n=3}^\infty \frac{{\varDelta_\mathrm{c}}^{n-2}}{n!}S_n < 1\,,
  \label{eq:4-09}
\end{equation}
otherwise, the mass fraction $\beta$ exceeds unity in the limit $\nu\gg1$. 
In this paper, we assume the above condition is satisfied.

When the values of reduced cumulants $S_n$ are of order unity, the above condition is safely satisfied. For example, when all the reduced cumulants $S_n$ have the same value, the condition~(\ref{eq:4-09}) with $\varDelta_\mathrm{c} \simeq 1/3$ implies $S_n \lesssim 8$. 
Thus it is not too restrictive. 
On the other hand, if the condition~(\ref{eq:4-09}) is not met, the asymptotic formula~(\ref{eq:4-06}) breaks down. 
This is because the resummation of leading contributions in the expansion in Eq.~(\ref{eq:4-03}) for the threshold model is not justified as $S_n$ are no longer of $\mathcal{O}(1)$. 
The same applies to the resummation in the peaks model, Eq.~(\ref{eq:a-15}). 
For example, when higher-order cumulants of a non-Gaussian model satisfy $\langle{\varDelta_R}^n\rangle_\mathrm{c} \sim \mathcal{O}(\sigma^n)$
[instead of $ \sim \mathcal{O}(\sigma^{2n-2})$, c.f.,
Eq.~(\ref{eq:4-01})], the resummation does not work
\cite{Riccardi:2021rlf}.

Comparing Eqs.~(\ref{eq:4-06}) and (\ref{eq:4-07}), and noting $\nu =
\varDelta_\mathrm{c}/\sigma$, we find that the effects of non-Gaussianity may be conveniently taken into account by replacing the threshold value for the PBH formation in the exponent of the Gaussian prediction, Eq.~(\ref{eq:4-07}), as 
\begin{equation}
  \varDelta_\mathrm{c} \rightarrow
  \varDelta_\mathrm{c}^\mathrm{eff} \equiv
  \varDelta_\mathrm{c}
  \sqrt{
    1- S    }\,;\quad
    S\equiv2\sum_{n=3}^\infty \frac{{\varDelta_\mathrm{c}}^{n-2}}{n!}S_n\,,
  \label{eq:4-10}
\end{equation}
apart from the prefactor $A(\nu)$. 
When $S$ is positive, the tail of the distribution function increases, which reduces the effective threshold in comparison with the Gaussian prediction, and the number of PBHs increases.
On the contrary, if $S$ is negative, the number of PBHs decreases. 
Thus, the expected number density of PBHs is exponentially sensitive to non-Gaussianity. 
We note that the effective threshold above,
$\varDelta_\mathrm{c}^\mathrm{eff}$, does not apply to the prefactor $A(\nu)$. Hence the effects of non-Gaussianity are not completely degenerate with the Gaussian case, though the change in the exponent dominates the effect on the number density of PBHs.

When the observational constraint is given by $\beta < \beta_0$, where $\beta_0\ll 1$ and thus $\ln(1/\beta_0)>0$, Eq.~(\ref{eq:4-06}) implies
\begin{equation}
  \sigma^2 < \frac{{\varDelta_\mathrm{c}}^2}{2\ln(1/\beta_0)}
  \left(
    1- S
  \right),
  \label{eq:4-20}
\end{equation}
where the logarithm of the prefactor $\ln A(\nu)$ is ignored, assuming $\ln(1/\beta_0) \gg \ln A(\nu)$. 
Therefore, an observational constraint
on the upper limit of the amplitude of primordial spectrum $\sqrt{\mathcal{P}_\calR}$ is tighter for $S>0$. 
If we only keep the leading term in $S$, $S\propto S_3$, this is in qualitative agreement with the results in Refs.~\cite{Byrnes:2012yx,Young:2015cyn} at linear order in $S_3$ (or in $f_{\rm NL}$ as discussed in the next section).
Nonlinear behaviors discussed in these references might be explained by the effect of ${f_\mathrm{NL}}^2$ in the variance,
$\sigma^2 \sim {\sigma_\mathrm{G}^2} + \mathrm{(const.)}\times
{f_\mathrm{NL}}^2{\sigma_\mathrm{G}^4}$. However, 
for a narrowly peaked power spectrum, the prefactor (const.) of this relation is suppressed by the width of the spectrum, $\epsilon=\Delta k/k_0\ll1$, where $k_0$ is the peak of the spectrum. 

As seen from Eq.~(\ref{eq:4-06}), the parameters of reduced cumulants,
$S_n$, with arbitrary higher orders may equally contribute if there
are of the same order of magnitude. In the following, however, we
mainly focus on the effects of bispectrum and trispectrum of the
primordial fluctuations, which are responsible to the skewness
parameter $S_3$ and the kurtosis parameter $S_4$. When the
higher-order cumulants $S_n$ with $n\geq 5$ are absent,
Eq.~(\ref{eq:4-06}) reduces to
\begin{equation}
  \beta \approx
  \beta^\mathrm{G}
  \exp\left[
    \nu^2 \varDelta_\mathrm{c}
    \left(\frac{S_3}{6}
      + \frac{{\varDelta_\mathrm{c}}S_4}{24}
      \right)
    \right].
  \label{eq:4-21}
\end{equation}
The condition of Eq.~(\ref{eq:4-09}) in this case is given by
\begin{equation}
  S_3 + \frac{S_4}{12} \lesssim 9,
  \label{eq:4-22}
\end{equation}
for $\varDelta_\mathrm{c} \simeq 1/3$. 
The skewness and kurtosis parameters, $S_3$ and $S_4$, are determined by non-Gaussian initial conditions. 
In the next section, we derive useful formulas to calculate these parameters in typical models
of primordial non-Gaussianity.

\section{\label{sec:Cumulants}
    Skewness and kurtosis in models of primordial non-Gaussianity
}

\subsection{Definitions}

The reduced cumulants of third and fourth orders, $S_3$ and $S_4$, are
called skewness and kurtosis parameters, respectively. They are
related to the bispectrum $B(\bm{k}_1,\bm{k}_2,\bm{k}_3)$ and the
trispectrum $T(\bm{k}_1,\bm{k}_2,\bm{k}_3,\bm{k}_4)$ of density
field $\varDelta$ by
\begin{align}
  S_3 &=
  \frac{1}{\sigma^4}
  \int_{\bm{k}_{123}=\bm{0}}
  B(\bm{k}_1,\bm{k}_2,\bm{k}_3)
  W(k_1R)W(k_2R)W(k_3R),
  \label{eq:5-01a}\\
  S_4 &=
  \frac{1}{\sigma^6}
  \int_{\bm{k}_{1234}=\bm{0}}
  T(\bm{k}_1,\bm{k}_2,\bm{k}_3,\bm{k}_4)
  W(k_1R)W(k_2R)W(k_3R)W(k_4R),
  \label{eq:5-01b}
\end{align}
where we use an abbreviated notation, $\bm{k}_{1\cdots n} \equiv
\bm{k}_1 + \cdots + \bm{k}_n$, and
\begin{equation}
  \int_{\bm{k}_{1\cdots n}=\bm{0}}\cdots
    \equiv
    \int \frac{d^3k_1}{(2\pi)^3}\cdots \frac{d^3k_n}{(2\pi)^3}
    (2\pi)^3 \delta_\mathrm{D}^3(\bm{k}_1 + \cdots + \bm{k}_n)\cdots.
  \label{eq:5-02}
\end{equation}
The bispectrum and trispectrum of the density field are related to those of the curvature perturbation by
\begin{align}
  B(\bm{k}_1,\bm{k}_2,\bm{k}_3)
  &= \mathcal{M}(k_1) \mathcal{M}(k_2) \mathcal{M}(k_3)
  B_\calR(\bm{k}_1,\bm{k}_2,\bm{k}_3),
  \label{eq:5-03a}\\
  T(\bm{k}_1,\bm{k}_2,\bm{k}_3,\bm{k}_4)
  &= \mathcal{M}(k_1) \mathcal{M}(k_2)
    \mathcal{M}(k_3) \mathcal{M}(k_4)
  T_\calR(\bm{k}_1,\bm{k}_2,\bm{k}_3,\bm{k}_4).
  \label{eq:5-03b}
\end{align}
Substituting Eqs.~(\ref{eq:2-04}), (\ref{eq:2-06}), (\ref{eq:5-03a}) and
(\ref{eq:5-03b}) into Eqs.~(\ref{eq:5-01a}) and (\ref{eq:5-01b}), we have
\begin{align}
  S_3
  &=
    \frac{1}{\sigma^4}
    \left(\frac{4}{9}\right)^3
    \int_{\bm{k}_{123}=\bm{0}}
    (k_1R)^2(k_2R)^2(k_3R)^2
    e^{-({k_1}^2+{k_2}^2+{k_3}^2)R^2/2}
    B_\calR(\bm{k}_1,\bm{k}_2,\bm{k}_3),
  \label{eq:5-04a}\\
  S_4
  &=
    \frac{1}{\sigma^6}
    \left(\frac{4}{9}\right)^4
    \int_{\bm{k}_{1234}=\bm{0}}
    (k_1R)^2(k_2R)^2(k_3R)^2(k_4R)^2
    e^{-({k_1}^2+{k_2}^2+{k_3}^2+{k_4}^2)R^2/2}
    T_\calR(\bm{k}_1,\bm{k}_2,\bm{k}_3,\bm{k}_4).
  \label{eq:5-04b}
\end{align}
The variance $\sigma^2$ is the same as ${\sigma_0}^2$ defined by
Eq.~(\ref{eq:3-05}). That is
\begin{equation}
  \sigma^2 = \left(\frac{4}{9}\right)^2
  \int \frac{k^2dk}{2\pi^2} (kR)^4 e^{-k^2R^2}
   P_\calR(k)\,,
  \label{eq:5-05}
\end{equation}
where $P_\calR(k)$ is the power spectrum of the comoving curvature perturbation.

\subsection{Skewness}

There are many models of primordial non-Gaussianity. One of the most
commonly assumed model for the bispectrum is the local-type
\cite{Gangui:1993tt,Verde:1999ij,Komatsu:2001rj}, which is given by
\begin{equation}
  B_\calR(\bm{k}_1,\bm{k}_2,\bm{k}_3) =
  \frac{6}{5} f_\mathrm{NL}
  \left[
    P_\calR(k_1) P_\calR(k_2) + \mathrm{cyc.}
  \right],
  \label{eq:5-14}
\end{equation}
where $f_\mathrm{NL}$ is a parameter of non-Gaussianity amplitude,
$+\,\mathrm{cyc.}$ represents the two terms obtained by cyclic
permutations of the preceding term with respect to $k_1$, $k_2$,
$k_3$. Among other types of non-Gaussianity for the bispectrum, popular alternatives are
the equilateral \cite{Creminelli:2005hu}, folded \cite{Meerburg:2009ys},
and orthogonal \cite{Senatore:2009gt} types.
These may be constructed from the following elements of the bispectrum:
\begin{equation}
  B^\mathrm{I}_{123}
  \equiv P_1 P_2 + \mathrm{cyc.},\quad
  B^\mathrm{II}_{123}
  \equiv (P_1 P_2 P_3)^{2/3},\quad
  B^\mathrm{III}_{123}
  \equiv {P_1}^{1/3} {P_2}^{2/3} P_3 + \mathrm{5\ perm.},
  \label{eq:5-15}
\end{equation}
where we denote $P_1 = P_\calR(k_1)$, $P_2 = P_\calR(k_2)$,
$P_3 = P_\calR(k_3)$ for simplicity, and $+\,\mathrm{5\ perm.}$
represents the 5 terms obtained by permutations of the preceding term. In terms of these elements,
the bispectrum for each type of non-Gaussianity is given by
\begin{align}
  B^\mathrm{loc}_\calR
  &=\frac{6}{5}f_\mathrm{NL}^\mathrm{loc}B^\mathrm{I}_{123}\,,
\label{eq:5-16-0}  \\
  B^\mathrm{eql}_\calR
  &= \frac{18}{5} f_\mathrm{NL}^\mathrm{eql}
    \left(
    - B^\mathrm{I}_{123} - 2B^\mathrm{II}_{123} + B^\mathrm{III}_{123}
    \right),
  \label{eq:5-16a}\\
  B^\mathrm{fol}_\calR
  &= \frac{18}{5} f_\mathrm{NL}^\mathrm{fol}
    \left(
    B^\mathrm{I}_{123} + 3B^\mathrm{II}_{123} - B^\mathrm{III}_{123}
    \right),
  \label{eq:5-16b}\\
  B^\mathrm{ort}_\calR
  &= \frac{18}{5} f_\mathrm{NL}^\mathrm{ort}
    \left(
    - 3B^\mathrm{I}_{123} - 8B^\mathrm{II}_{123} + 3B^\mathrm{III}_{123}
    \right).
  \label{eq:5-16c}
\end{align}

We introduce the skewness parameter elements corresponding to the
above three elements $B^A_{123}$ ($A={\rm I},{\rm II},{\rm III}$) of
the bispectrum as
\begin{equation}
  S^A_3
  \equiv
    \frac{1}{\sigma^4}
    \left(\frac{4}{9}\right)^3
    \int_{\bm{k}_{123}=\bm{0}}
    (k_1R)^2 (k_2R)^2 (k_3R)^2
    W(k_1R)W(k_2R)W(k_3R)B^A_{123}\,.
  \label{eq:5-17}
\end{equation}
The skewness parameters for local, equilateral, folded, and orthogonal
types, which we denote by $S^\mathrm{loc}_3$, $S^\mathrm{eql}_3$,
$S^\mathrm{fol}_3$, $S^\mathrm{ort}_3$, respectively, are given by
linear superpositions of $S^A_3$ ($A={\rm I},{\rm II},{\rm III}$) in
exactly the same forms as Eqs.~(\ref{eq:5-16-0}) -- (\ref{eq:5-16c}),
\begin{align}
  S^\mathrm{loc}_3
  &= \frac{6}{5} f_\mathrm{NL}^\mathrm{loc} S^\mathrm{I}_3,
  \label{eq:5-18a}\\
  S^\mathrm{eql}_3
  &= \frac{18}{5} f_\mathrm{NL}^\mathrm{eql}
    \left(
    - S^\mathrm{I}_3 - 2S^\mathrm{II}_3 + S^\mathrm{III}_3
    \right),
  \label{eq:5-18b}\\
  S^\mathrm{fol}_3
  &= \frac{18}{5} f_\mathrm{NL}^\mathrm{fol}
    \left(
    S^\mathrm{I}_3 + 3S^\mathrm{II}_3 - S^\mathrm{III}_3
    \right),
  \label{eq:5-18c}\\
  S^\mathrm{ort}_3
  &= \frac{18}{5} f_\mathrm{NL}^\mathrm{ort}
    \left(
    - 3S^\mathrm{I}_3 - 8S^\mathrm{II}_3 + 3S^\mathrm{III}_3
    \right).
  \label{eq:5-18d}
\end{align}
We note that, excluding the signs of $f_{\rm NL}^X$ ($X=\mathrm{loc}, \mathrm{eql}, \mathrm{fol}, \mathrm{ort}$) in the coefficients, $S^\mathrm{\rm loc}_3$ is positive definite, while the sign of the other two is indeterminate.
Given the values of $f_\mathrm{NL}^{X}$, these skewness parameters are uniquely determined once the primordial power spectrum $P_\calR(k)$ is known. 
In the high-peaks limit, the effects of skewness in each non-Gaussian type on the abundance of PBHs are given by substituting the results into Eq.~(\ref{eq:4-21}).

The skewness parameter elements are given by substituting Eqs.~(\ref{eq:5-15}) into Eq.~(\ref{eq:5-17}). 
To represent the results in a convenient, compact form, we introduce the dimensionless curvature perturbation power spectrum,
\begin{equation}
  \mathcal{P}_\calR(k) \equiv \frac{k^3P_\calR(k)}{2\pi^2}\,.
  \label{eq:5-19}
\end{equation}
Changing the integration variables as $\bm{p}=\bm{k}_1R$,
$\bm{q}=\bm{k}_2R$ and $r=|\bm{p}+\bm{q}|$, where the variable $r$ describes the angular degrees of freedom, $\mu = \bm{p}\cdot\bm{q}/(pq) = (r^2 - p^2 - q^2)/(2pq)$,
some of the angular integrations may be analytically calculated to give
\begin{align}
  S^\mathrm{I}_3
  &= \frac{3}{\sigma^4} \left(\frac{4}{9}\right)^3
    \int_0^\infty dp\,dq\,e^{-p^2-q^2} pq
    \left[(p^2+q^2+2)\frac{\sinh(pq)}{pq} - 2\cosh(pq)\right]
    \mathcal{P}_\calR\left(\frac{p}{R}\right)
    \mathcal{P}_\calR\left(\frac{q}{R}\right),
    \label{eq:5-20a}\\
  S^\mathrm{II}_3
  &= \frac{1}{\sigma^4} \left(\frac{4}{9}\right)^3
    \int_0^\infty dp\,dq\,e^{-(p^2+q^2)/2} pq
    \left[
    \mathcal{P}_\calR\left(\frac{p}{R}\right)
    \mathcal{P}_\calR\left(\frac{q}{R}\right)
    \right]^{2/3}
    \int_{|p-q|}^{p+q} \frac{dr}{2}\,r\,e^{-r^2/2}
    \left[
    \mathcal{P}_\calR\left(\frac{r}{R}\right)
    \right]^{2/3},
    \label{eq:5-20b}\\
  S^\mathrm{III}_3
  &= \frac{6}{\sigma^4} \left(\frac{4}{9}\right)^3
    \int_0^\infty dp\,dq\,p^2q\,e^{-(p^2+q^2)/2}
    \left[\mathcal{P}_\calR\left(\frac{p}{R}\right)\right]^{1/3}
    \left[\mathcal{P}_\calR\left(\frac{q}{R}\right) \right]^{2/3}
    \int_{|p-q|}^{p+q} \frac{dr}{2}\,e^{-r^2/2}\,
    \mathcal{P}_\calR\left(\frac{r}{R}\right).
    \label{eq:5-20c}
\end{align}
Thus, all the skewness parameters in the three types of non-Gaussianity can be computed from the above equations for an arbitrary power spectrum $P_\calR(k)$. 
The variance $\sigma^2$ of Eq.~(\ref{eq:5-05}) is similarly given by a one-dimensional integral:
\begin{equation}
  \sigma^2
  = \left(\frac{4}{9}\right)^2
  \int dp\,p^3 e^{-p^2}
  \mathcal{P}_\calR\left(\frac{p}{R}\right).
  \label{eq:5-21}
\end{equation}

\subsection{Kurtosis}

In contrast to the case of the bispectrum, not so many variations in the types of the primordial trispectrum
have been proposed. Lacking considerations on the general types of the trispectrum, here we focus on the most popular type, that is the generalized local-type \cite{Byrnes:2006vq},
\begin{equation}
  T_\calR(\bm{k}_1,\bm{k}_2,\bm{k}_3,\bm{k}_4) =
  \frac{54}{25} g_\mathrm{NL}
  \left[P_1 P_2 P_3 + \mathrm{3\ perm.}\right]
  + \tau_\mathrm{NL}
  \left[P_1 P_2 P_{23} + \mathrm{11\ perm.}\right],
  \label{eq:5-30}
\end{equation}
where $P_{23}=P_\calR(|\bm{k}_2+\bm{k}_3|)$, with
$g_\mathrm{NL}$ and $\tau_\mathrm{NL}$ being the parameters.
If the primordial perturbations emerge from the quantum fluctuations of a
single scalar field, the last parameter is related to the parameter of
the local-type bispectrum by
$\tau_\mathrm{NL} = (36/25){f_\mathrm{NL}}^2$ \cite{Boubekeur:2005fj}.
If multiple scalar fields are involved, there is an inequality,
$\tau_\mathrm{NL} > (36/25){f_\mathrm{NL}}^2$ \cite{Suyama:2007bg}.
In this model, we define elements of the trispectrum by
\begin{equation}
  T^\mathrm{I}_{1234}
  \equiv P_1 P_2 P_3 + \mathrm{3\ perm.},\quad
  T^\mathrm{II}_{1234}
  \equiv P_1 P_2 P_{23} + \mathrm{11\ perm.}.
  \label{eq:5-31}
\end{equation}
The corresponding elements of kurtosis are given by
\begin{equation}
  S^A_4
  \equiv
    \frac{1}{\sigma^6}
    \left(\frac{4}{9}\right)^4
    \int_{\bm{k}_{1234}=\bm{0}}
    (k_1R)^2 (k_2R)^2 (k_3R)^2 (k_4R)^2
    W(k_1R)W(k_2R)W(k_3R)W(k_4R)T^A_{1234},
  \label{eq:5-32}
\end{equation}
where $A=\mathrm{I}, \mathrm{II}$. We note that both elements are positive definite, as clear from their definitions. 
The resulting kurtosis for the local-type is given by
\begin{equation}
  S_4 =
  \frac{54}{25} g_\mathrm{NL} S^\mathrm{I}_4
  + \tau_\mathrm{NL} S^\mathrm{II}_4.
  \label{eq:5-33}
\end{equation}
Once the primordial power spectrum is given, one can evaluate the kurtosis. 
The effect on the abundance of PBHs in the high-peaks limit is given by substituting the result into Eq.~(\ref{eq:4-21}).

The kurtosis parameter elements are given by substituting Eqs.~(\ref{eq:5-31}) into Eq.~(\ref{eq:5-32}).
Changing integration variables as
$\bm{p}=\bm{k}_1R$, $\bm{q}=\bm{k}_2R$, $\bm{r} =(\bm{k}_2+\bm{k}_3)R$, $\mu = \bm{p}\cdot\bm{r}/(pr)$, $\mu' = - \bm{q}\cdot\bm{r}/(qr)$, and expressing $\mu'$ in terms of $s=|\bm{q}-\bm{r}|$, some of the angular integrations can be analytically calculated (see Ref.~\cite{Matsubara:2020knr} for the same type of calculation). The results are
\begin{align}
  S^\mathrm{I}_4
  &= \frac{4}{\sigma^6} \left(\frac{4}{9}\right)^4
    \int_0^\infty dp\,dq\,dr\,
    e^{-p^2}e^{-(q^2+r^2)/2} pr
    \nonumber\\
  &\hspace{3.5pc}
    \times
    \left[(p^2 + r^2 + 2)\frac{\sinh(pr)}{pr} - 2\cosh(pr)\right]
    \mathcal{P}_\calR\left(\frac{p}{R}\right)
    \mathcal{P}_\calR\left(\frac{q}{R}\right)
    \int_{|q-r|}^{q+r}\frac{ds}{2} e^{-s^2/2}
    \mathcal{P}_\calR\left(\frac{s}{R}\right),
    \label{eq:5-34a}\\
  S^\mathrm{II}_4
  &= \frac{12}{\sigma^6} \left(\frac{4}{9}\right)^4
    \int_0^\infty dp\,dq\,dr\,e^{-p^2-q^2-r^2}\frac{pq}{r}
    \left[(p^2+r^2+2)\frac{\sinh(pr)}{pr} - 2\cosh(pr)\right]
    \nonumber\\
  &\hspace{9pc}
    \times
    \left[(q^2+r^2+2)\frac{\sinh(qr)}{qr} - 2\cosh(qr)\right]
    \mathcal{P}_\calR\left(\frac{p}{R}\right)
    \mathcal{P}_\calR\left(\frac{q}{R}\right)
    \mathcal{P}_\calR\left(\frac{r}{R}\right).
    \label{eq:5-34b}
\end{align}
The above formulas enable us to compute all the kurtosis parameters of the local type for an arbitrary primordial power spectrum $P_\calR(k)$.

\section{\label{sec:Narrow}
  Power spectrum with a narrow peak}

In the previous section, we derived formulas for the skewness and kurtosis that can be used to evaluate their effects on the abundance of PBHs for a general primordial power spectrum.
In this section, we focus on the case of a narrowly peaked spectrum, which would lead to a nearly monochromatic mass function of PBHs.

We consider the case that the primordial power spectrum peaked at a wavenumber $k_0$. When the sharpness of the peak is extreme, we can approximately substitute
\begin{equation}
  f(k) \mathcal{P}_\calR(k) \rightarrow f(k_0) \mathcal{P}_\calR(k),
  \label{eq:6-01}
\end{equation}
where $f(k)$ is an arbitrary function of $k$. Applying this
substitution, the variance of Eq.~(\ref{eq:5-21}) reduces to
\begin{equation}
  \sigma^2 \simeq \left(\frac{4}{9}\right)^2
  e^{-{k_0}^2R^2}(k_0R)^4 \int \frac{dk}{k}\,
  \mathcal{P}_\calR(k)\,,
  \label{eq:6-02}
\end{equation}
and the skewness parameter elements in Eqs.~(\ref{eq:5-20a}) -- (\ref{eq:5-20c}) reduce to 
\begin{align}
  S^\mathrm{I}_3
  &\simeq
    \frac{27}{2}\,
    \frac{1}{(k_0R)^4}
    \left[({k_0}^2R^2+1) \frac{\sinh({k_0}^2R^2)}{{k_0}^2R^2} -
    \cosh({k_0}^2R^2)\right]\,,
    \label{eq:6-02a}\\
  S^\mathrm{II}_3
  &\simeq \frac{9}{8}\,
    \frac{e^{{k_0}^2R^2/2}}{{k_0}^2R^2}
    \frac{{C_{2/3}}^3}{{C_1}^2},
    \label{eq:6-02b}\\
  S^\mathrm{III}_3
  &\simeq \frac{27}{4}\,
    \frac{e^{{k_0}^2R^2/2}}{{k_0}^2R^2}
    \frac{C_{1/3}C_{2/3}}{C_1},
    \label{eq:6-02c}
\end{align}
where we have introduced
\begin{equation}
  C_\alpha \equiv
  \int \frac{dk}{k} \left[\mathcal{P}_\calR(k)\right]^\alpha.
  \label{eq:6-03}
\end{equation}
The integral range of the above is localized in the vicinity of $k_0$.
We note that $S^\mathrm{I}_3$ takes the minimum value
$S^\mathrm{I}_{3}|_{\rm min} \approx4.93$ at $k_0R\approx1.7$. We also
note that $S^\mathrm{II}_3$ and $S^\mathrm{III}_3$ depend on the
detailed shape of the peak in the power spectrum through the integrals
$C_\alpha$.

Due to the assumption of a narrow peak, the values of
integrals $C_\alpha$ are small. If we characterize the
narrowness of the power spectrum by $\epsilon$ in the space of
wavenumber, we have $C_\alpha \sim \mathcal{O}(\epsilon)$
for $\alpha > 0$. Thus the parameters $S^\mathrm{II}_3$ and
$S^\mathrm{III}_3$ are of order $\sim \epsilon\,S^\mathrm{I}_3$,
\begin{equation}
  S^\mathrm{I}_3 \gg S^\mathrm{II}_3,  S^\mathrm{III}_3.
  \label{eq:6-04}
\end{equation}
This is a remarkable property of a narrow power spectrum that
simplifies the bispectrum in Eqs.~(\ref{eq:5-16-0}) --
(\ref{eq:5-16c}). For example, if the narrow shape of the power
spectrum is given by a rectangular function of width $k_0 \epsilon$,
\begin{equation}
  \mathcal{P}_\calR(k) = 
  \begin{cases}
    A_0,
    & \displaystyle
    \mathrm{if\ \ } 1 - \frac{\epsilon}{2}
    \leq \frac{k}{k_0} \leq 1 + \frac{\epsilon}{2},
    \\
    0 & \mathrm{otherwise},
    \end{cases}
    \label{eq:6-05}
\end{equation}
where $A_0$ is a normalization constant, then we have
$C_\alpha = {A_0}^\alpha \epsilon$, which gives
\begin{equation}
  S^\mathrm{II}_3
  \simeq \frac{9\epsilon}{8}
    \frac{e^{{k_0}^2R^2/2}}{(k_0R)^2}, \qquad
  S^\mathrm{III}_3
  \simeq \frac{27\epsilon}{4}
    \frac{e^{{k_0}^2R^2/2}}{(k_0R)^2}.
  \label{eq:6-06}
\end{equation}
If the narrow shape is given by a Gaussian function of the full width
at half maximum (FWHM)\footnote{FWHM for a normal distribution of
  standard deviation $\sigma$ is given by
  $2\sqrt{2\ln 2}\,\sigma \simeq 2.35482 \sigma$.} $k_0\epsilon$,
\begin{equation}
  \mathcal{P}_\calR(k)
  = A_0 \exp\left[-\frac{4\ln 2}{k_0^2\epsilon^2} (k-k_0)^2\right],
  \label{eq:6-07}
\end{equation}
then we have
$C_\alpha = \sqrt{\pi/\ln 2}\,{\alpha}^{-1/2}{A_0}^\alpha\epsilon$, which gives
\begin{equation}
  S^\mathrm{II}_3 \simeq \frac{81\pi^{1/2}\epsilon}{32\sqrt{6\ln 2}}
  \frac{e^{{k_0}^2R^2/2}}{(k_0R)^2}, \qquad
  S^\mathrm{III}_3 \simeq \frac{81\pi^{1/2}\epsilon}{8\sqrt{2\ln 2}}
  \frac{e^{{k_0}^2R^2/2}}{(k_0R)^2}.
\label{eq:6-08}
\end{equation}
In any case, these parameters are much smaller than $S^\mathrm{I}_3$
by a factor of $\epsilon$.

Due to this property, the skewness parameters in Eqs.~(\ref{eq:5-18a})
-- (\ref{eq:5-18d}) are solely determined by a single element
$S^\mathrm{I}_3$. Taking account of contributions from all types,
the skewness of the density field is given by
\begin{equation}
  S_3 \simeq \frac{6}{5}
  \left(
    f_\mathrm{NL}^\mathrm{loc}
    - 3 f_\mathrm{NL}^\mathrm{eql}
    + 3 f_\mathrm{NL}^\mathrm{fol}
    -9 f_\mathrm{NL}^\mathrm{ort}
  \right) S^\mathrm{I}_3.
  \label{eq:6-10}
\end{equation}
Thus different types of non-Gaussianity give the same form of skewness
with different amplitudes. This means that one cannot distinguish
non-Gaussian types only with the amplitude of the skewness. Apart from
the numerical coefficients and the signs in (\ref{eq:6-10}), this fact
may be considered rather trivial because the abundance of PBHs is just
a single number. Different non-Gaussian types cannot be distinguished
by a single number.

Similarly, the substitution of Eq.~(\ref{eq:6-01}) in
Eqs.~(\ref{eq:5-34a}) and (\ref{eq:5-34b}), the kurtosis parameter
elements are given by
\begin{align}
  S^\mathrm{I}_4
  &\simeq
  \frac{243}{8} \frac{1}{(k_0R)^8}
    \Biggl\{
    - e^{-{k_0}^2R^2} \sinh\left(2{k_0}^2R^2\right)
    + \frac{e^{3{k_0}^2R^2/2}}{4k_0R}
    \sqrt{2\pi}
    \left[
    2
    -3\, \mathrm{erfc}
    \left(\frac{k_0R}{\sqrt{2}}\right)
     + \mathrm{erfc}
    \left(\frac{3k_0R}{\sqrt{2}}\right) 
    \right]
    \Biggr\},
    \label{eq:6-11a}\\
  S^\mathrm{II}_4
  &\simeq
    \frac{243}{(k_0R)^8}
    \left[({k_0}^2R^2+1) \frac{\sinh({k_0}^2R^2)}{{k_0}^2R^2} -
    \cosh({k_0}^2R^2)\right]^2
  \simeq
    \frac{4}{3}
    \left(S^\mathrm{I}_3\right)^2.
    \label{eq:6-11b}
\end{align}
The narrowness parameter $\epsilon$ discussed above cancels in this case, and therefore $S^\mathrm{I}_4$ and $S^\mathrm{II}_4$ are equally of order unity in $\epsilon$.
Hence both parameters do not depend on the detailed shape of the peak, as in the case of $S^\mathrm{I}_3$.

Taking into account all the leading contributions to the skewness and
kurtosis, the mass fraction with non-Gaussianity given by
Eq.~(\ref{eq:4-21}) in the high peaks limit with a narrow spectrum may be expressed in a single formula,
\begin{equation}
  \beta \simeq \beta^\mathrm{G}
  \exp\left\{
    \nu^2
      \varDelta_\mathrm{c}
      \left[
        \frac{1}{5} f_\mathrm{NL}^\mathrm{eff} S^\mathrm{I}_3
      + 
      \frac{{\varDelta_\mathrm{c}}}{12}
      \left(
        \frac{27}{25} g_\mathrm{NL} S^\mathrm{I}_4
        + \frac{2}{3} \tau_\mathrm{NL} (S^\mathrm{I}_3)^2
      \right)
    \right]
  \right\},
  \label{eq:6-12}
\end{equation}
where
\begin{equation}
  f_\mathrm{NL}^\mathrm{eff} \equiv
  f_\mathrm{NL}^\mathrm{loc}
  - 3 f_\mathrm{NL}^\mathrm{eql}
  + 3 f_\mathrm{NL}^\mathrm{fol}
  -9 f_\mathrm{NL}^\mathrm{ort}\,,
  \label{eq:6-13}
\end{equation}
irrespectively of the detailed shape of the narrow peak. 
The evaluation of the above formula requires only
two parameters of skewness and kurtosis, $S^\mathrm{I}_3$ and
$S^\mathrm{I}_4$.

For the Gaussian part of the mass fraction, $\beta^\mathrm{G}$, the peaks model, Eq.~(\ref{eq:4-08}), contains a factor
$R\sigma_1/\sigma$. Taking the limit of a narrow spectrum in
Eq.~(\ref{eq:3-05}), we obtain $\sigma_j = {k_0}^j \sigma$. Therefore, we have
\begin{equation}
  \beta^\mathrm{G} = 
  \frac{e^{-\nu^2/2}}{\sqrt{2\pi}} \times
  \begin{cases}
    \displaystyle
    \nu^{-1}, & \mathrm{(threshold \ model)}, \\
    \displaystyle
    \left(\frac{k_0R}{\sqrt{3}}\right)^3(\nu^2-1),
    & \mathrm{(peaks\ model)},
  \end{cases}
  \label{eq:6-14}
\end{equation}
in the limit of a narrow spectrum. 

\section{\label{sec:Numrical}
  Numerical demonstrations}

In this section, to obtain an intuitive sense of our results, and to find the range of validity of the narrow-peak limit approximation of the spectrum, we numerically evaluate the formulas we derived in the previous section.

\begin{figure}[tbp]
\centering
\includegraphics[width=.48\textwidth]{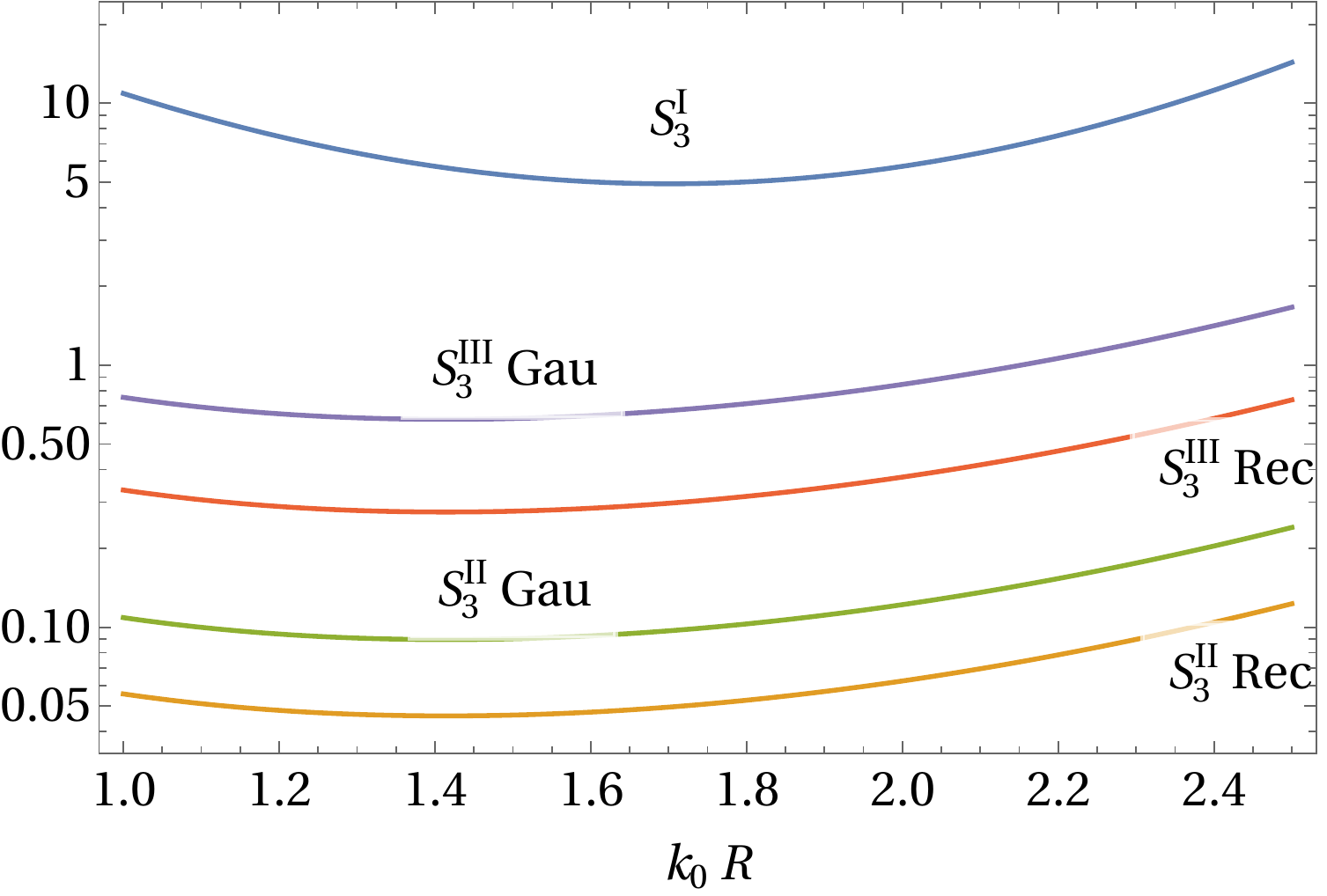}
\hfill
\includegraphics[width=.48\textwidth]{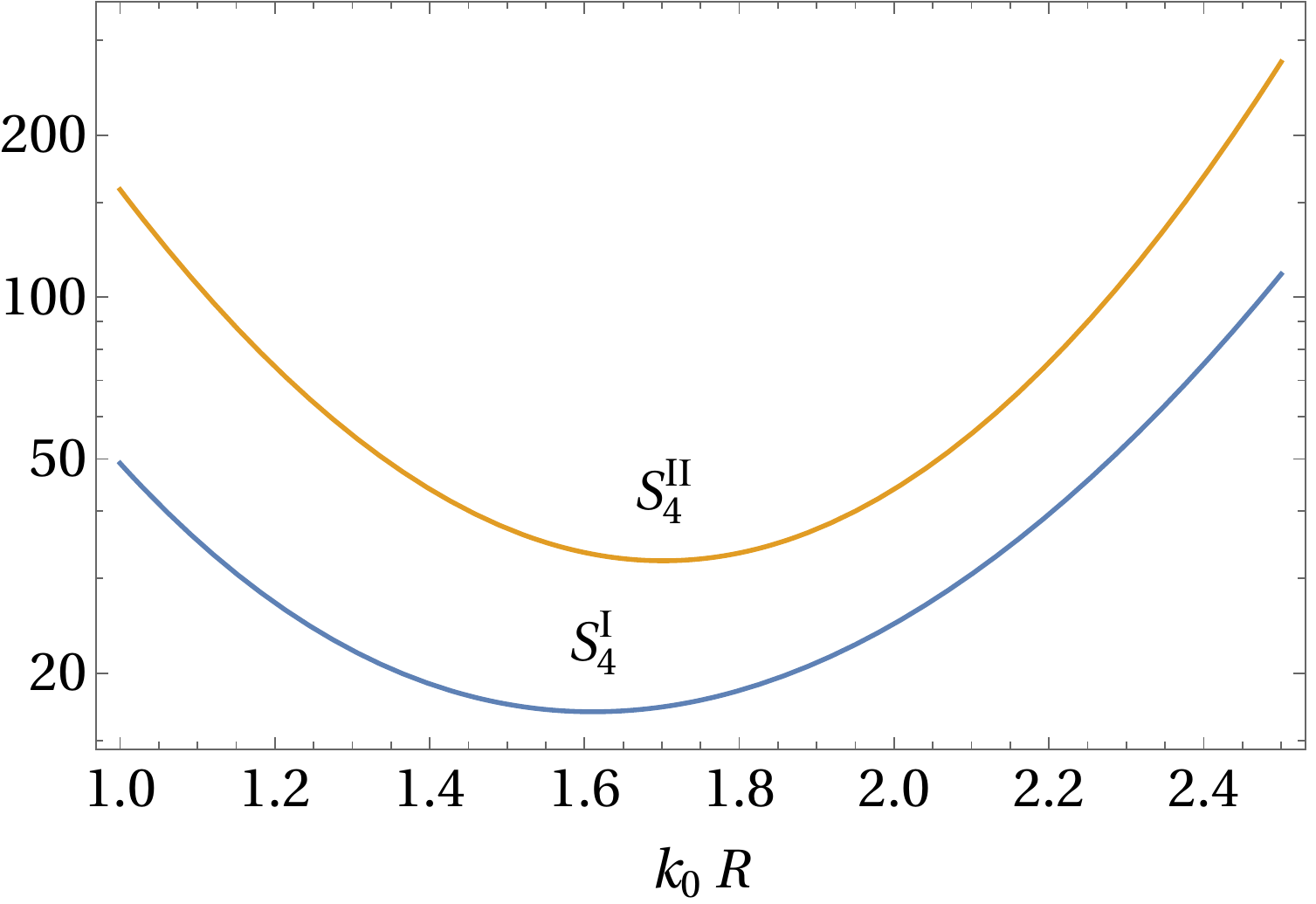}
\caption{\label{fig:01} Left panel: Elements of skewness parameters as
  a function of $k_0R$ in the narrow limit of spectral shapes. The
  elements $S_3^\mathrm{I}$, $S_3^\mathrm{II}$ and $S_3^\mathrm{III}$
  are plotted. For the latter two, they depend on the shape of the
  narrow power spectra. Rectangular shapes are indicated by ``Rec''
  and Gaussian shapes are indicated by ``Gau'' in the labels of the
  curves, and the narrowness parameter is taken to be $\epsilon = 0.03$ in
  the plot as an example. For another value of the narrowness
  parameter, $S_3^\mathrm{II}$ and $S_3^\mathrm{III}$ are simply
  proportional to $\epsilon$ in the plot. Right panel: Elements of
  kurtosis parameters in the narrow limit of spectral shapes. The
  elements $S_4^\mathrm{I}$ and $S_4^\mathrm{II}$ are plotted. These
  parameters do not depend on the details of the shapes of the narrow spectra.
}
\end{figure}

In the left panel of Fig.~\ref{fig:01}, the skewness
parameter elements, $S_3^\mathrm{I}$, $S_3^\mathrm{II}$ and $S_3^\mathrm{III}$ computed from Eqs.~(\ref{eq:6-02a}) -- (\ref{eq:6-02c}) in the narrow peak limit are shown as functions of $k_0R$. 
The element $S_3^\mathrm{I}$ does not depend on the precise shape of the spectrum in the narrow limit, while the other two elements, $S_3^\mathrm{II}$ and $S_3^\mathrm{III}$, depend on the shape of the narrow peak, hence the rectangular case, Eq.~(\ref{eq:6-05}) and the Gaussian case, Eq.~(\ref{eq:6-07}) are presented. 
The width of the peak characterized by the narrowness parameter $\epsilon$ is taken to be $\epsilon = 0.03$ in the plot just for illustration. We recall that the elements $S_3^\mathrm{II}$ and $S_3^\mathrm{III}$ are simply proportional to $\epsilon$. 
As seen from Fig.~\ref{fig:01}, $S_3^\mathrm{I}$ dominates the others when $\epsilon$ is small. For the value of $\epsilon = 0.03$, $S_3^\mathrm{I}$ is approximately an order of magnitude larger than the other two.

In the right panel of Fig.~\ref{fig:01}, the kurtosis
parameter elements, $S_4^\mathrm{I}$ and $S_4^\mathrm{II}$ computed by using Eqs.~(\ref{eq:5-34a}) and (\ref{eq:5-34b}) are shown in the narrow peak limit. Both of these two do not depend on the precise shape of the narrowness parameter $\epsilon$, and both contribute to the kurtosis parameter, irrespective of the narrowness of the spectrum. 
Since the values of the two elements $S_4^\mathrm{I}$ and $S_4^\mathrm{II}$ are roughly of $\mathcal{O}(10^1)$ - $\mathcal{O}(10^2)$, Eq.~(\ref{eq:4-22}) requires that the parameters $g_\mathrm{NL}$ and $\tau_\mathrm{NL}$ should be somewhat smaller than unity to satisfy the condition for the consistency of the approximation.

\begin{figure}[tbp]
\centering
\includegraphics[width=.48\textwidth]{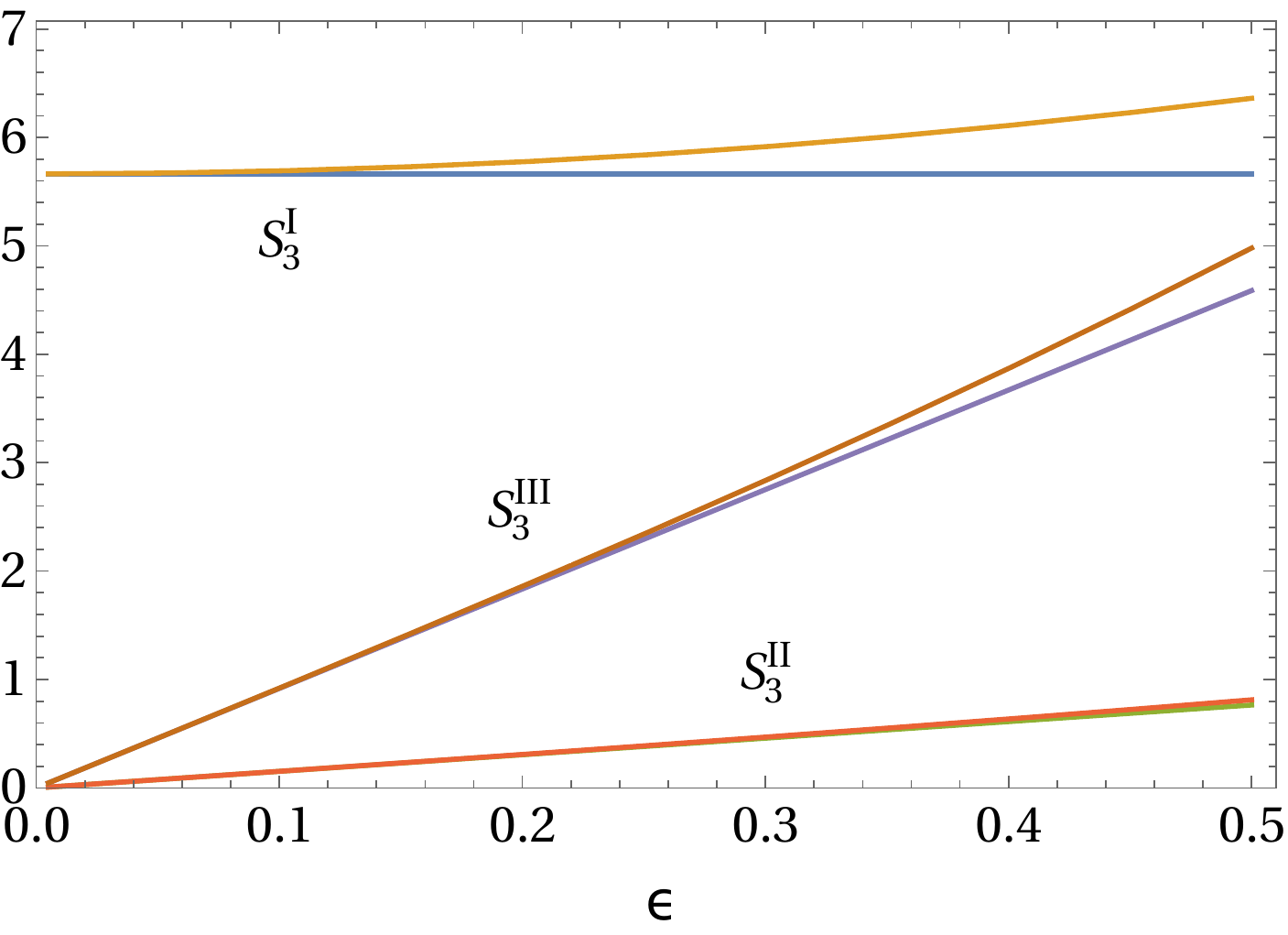}
\hfill
\includegraphics[width=.48\textwidth]{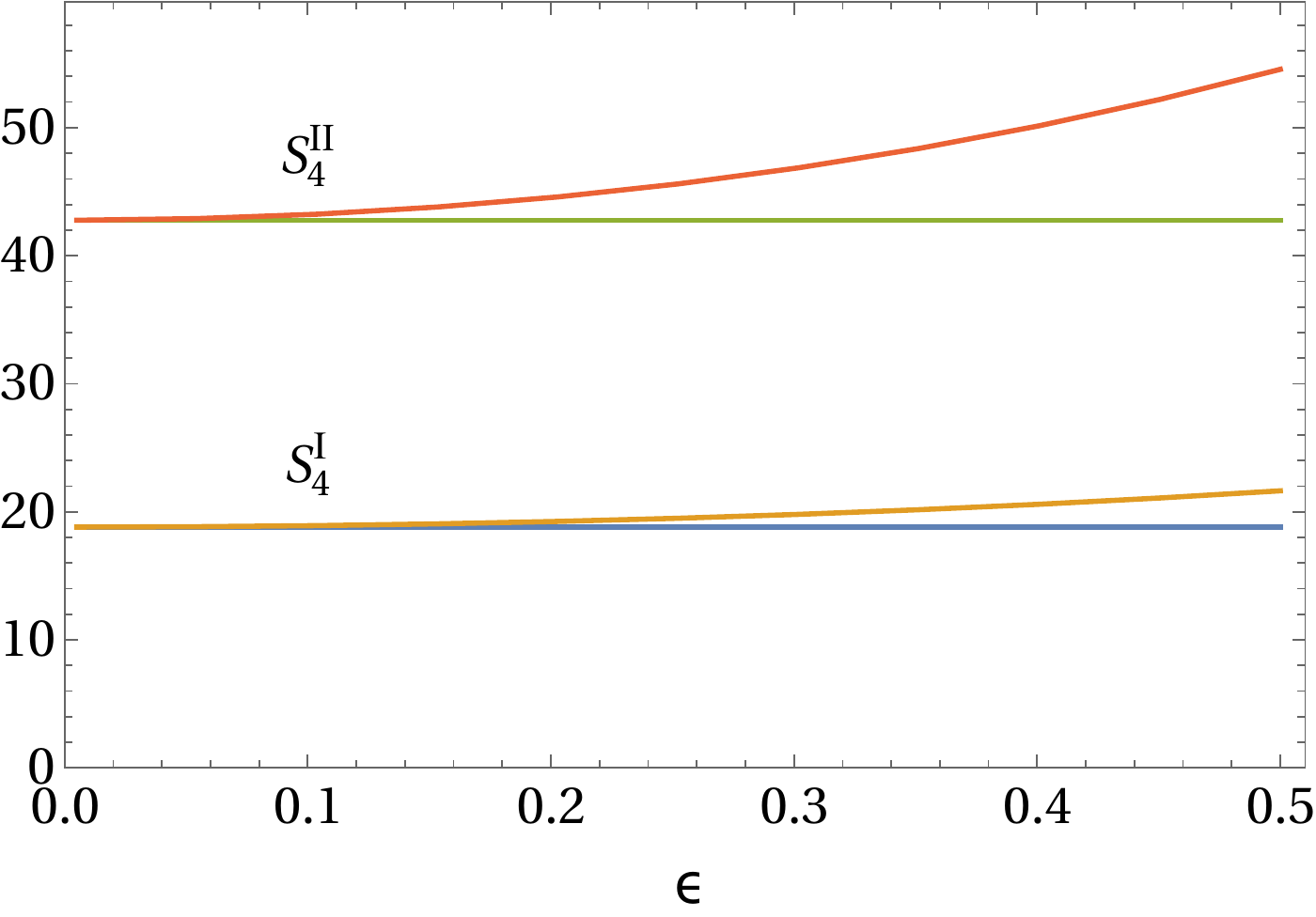}
\caption{\label{fig:02} Left panel: Comparisons between exact
  integrations and approximations of the narrow limit of primordial
  spectrum for three elements of the skewness parameter,
  $S_3^\mathrm{I}$, $S_3^\mathrm{II}$ and $S_3^\mathrm{III}$, as
  functions of the narrowness parameter $\epsilon$. The spectral shape
  is assumed to have the rectangular shape, and $k_0R = \sqrt{2}$. In
  the limits that the narrowness parameter $\epsilon$ is small, the
  approximation of the narrow limit and the results of exact
  integrations converge to the same lines to the left. The lines of
  the narrow limit are given by straight lines as the narrow limit of
  $S_3^\mathrm{I}$ is constant and those of the other are linearly
  proportional to $\epsilon$. Curved lines above the straight lines
  correspond to the results of exact integrations. Right panel: Same
  as the left panel, but elements of skewness parameter are compared. }
\end{figure}

In the left panel of Fig.~\ref{fig:02}, 
comparisons of the narrow limit approximation with the exact numerical results are made for the skewness parameter elements, $S_3^\mathrm{I}$, $S_3^\mathrm{II}$ and $S_3^\mathrm{III}$, and for the kurtosis parameter elements, $S_4^\mathrm{I}$ and $S_4^\mathrm{II}$. 
The right panel shows the comparison of the exact integrations of Eqs.~(\ref{eq:5-20a}) -- (\ref{eq:5-20c}) with the narrow limit approximation for the elements of skewness. 
The narrow limit of $S_3^\mathrm{I}$ is independent of $\epsilon$, while those of $S_3^\mathrm{II}$ and $S_3^\mathrm{III}$ are linearly proportional to $\epsilon$. 
The curved lines represent the exact numerical results, and they converge to the corresponding results in the narrow limit approximation as $\epsilon\to0$. 
As can be seen from the plot, the approximation of the narrow limit is accurate for $\epsilon \lesssim 0.3$. 
For larger values of $\epsilon$, the approximations are still fairly good, although the value of $S_3^\mathrm{III}$ is non-negligible in comparison to that of $S_3^\mathrm{I}$. 

In the right panel of Fig.~\ref{fig:02}, the kurtosis parameter elements are similarly compared. The narrow limits are given by the constant lines, while the exact numerical results are given by the curved lines. The narrow limit approximation is also valid for sufficiently small $\epsilon$, though the range of validity seems slightly smaller, $\epsilon\lesssim0.2$ in comparison with the case of the skewness.

\begin{figure}[tbp]
\centering
\includegraphics[width=.48\textwidth]{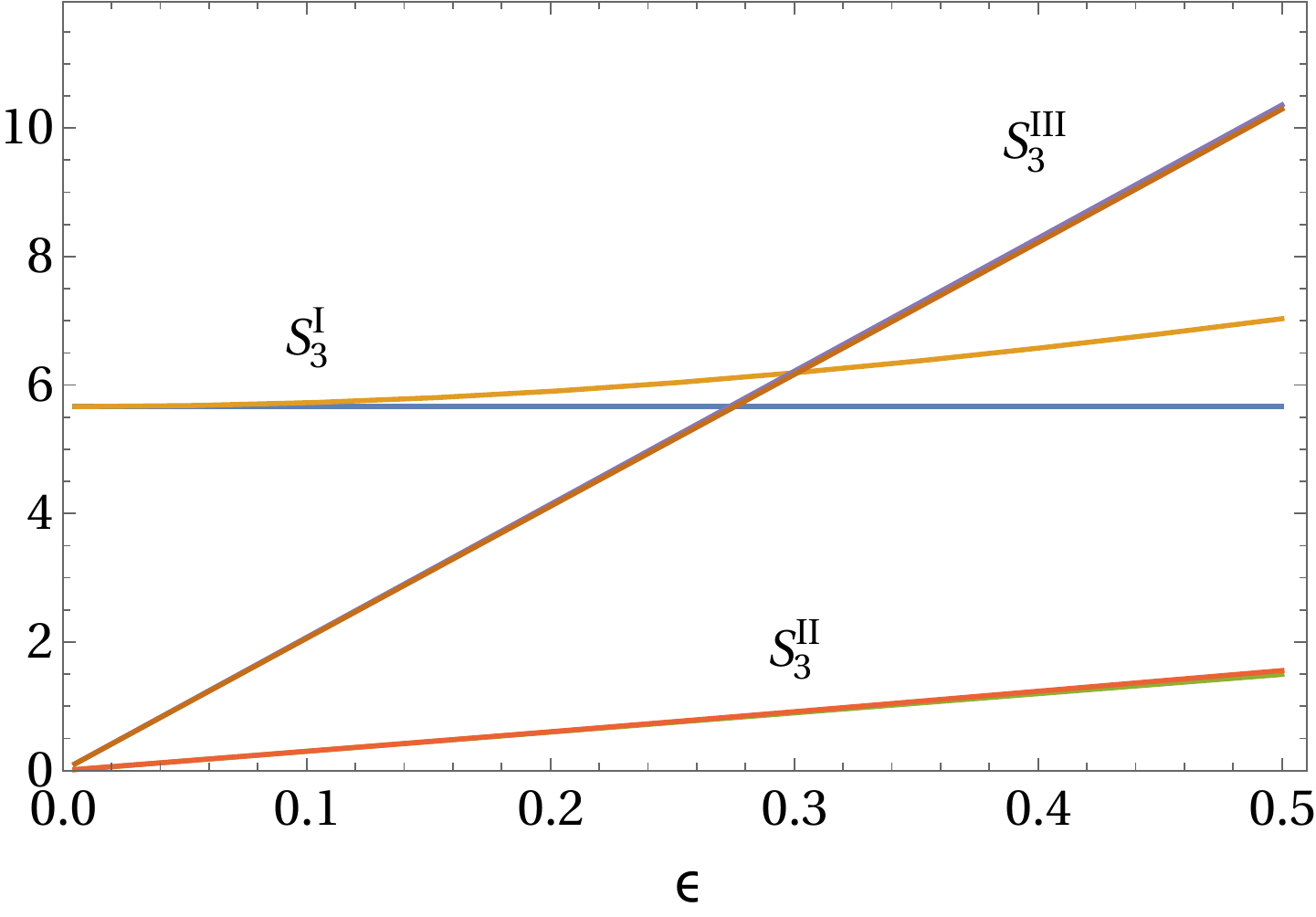}
\hfill
\includegraphics[width=.48\textwidth]{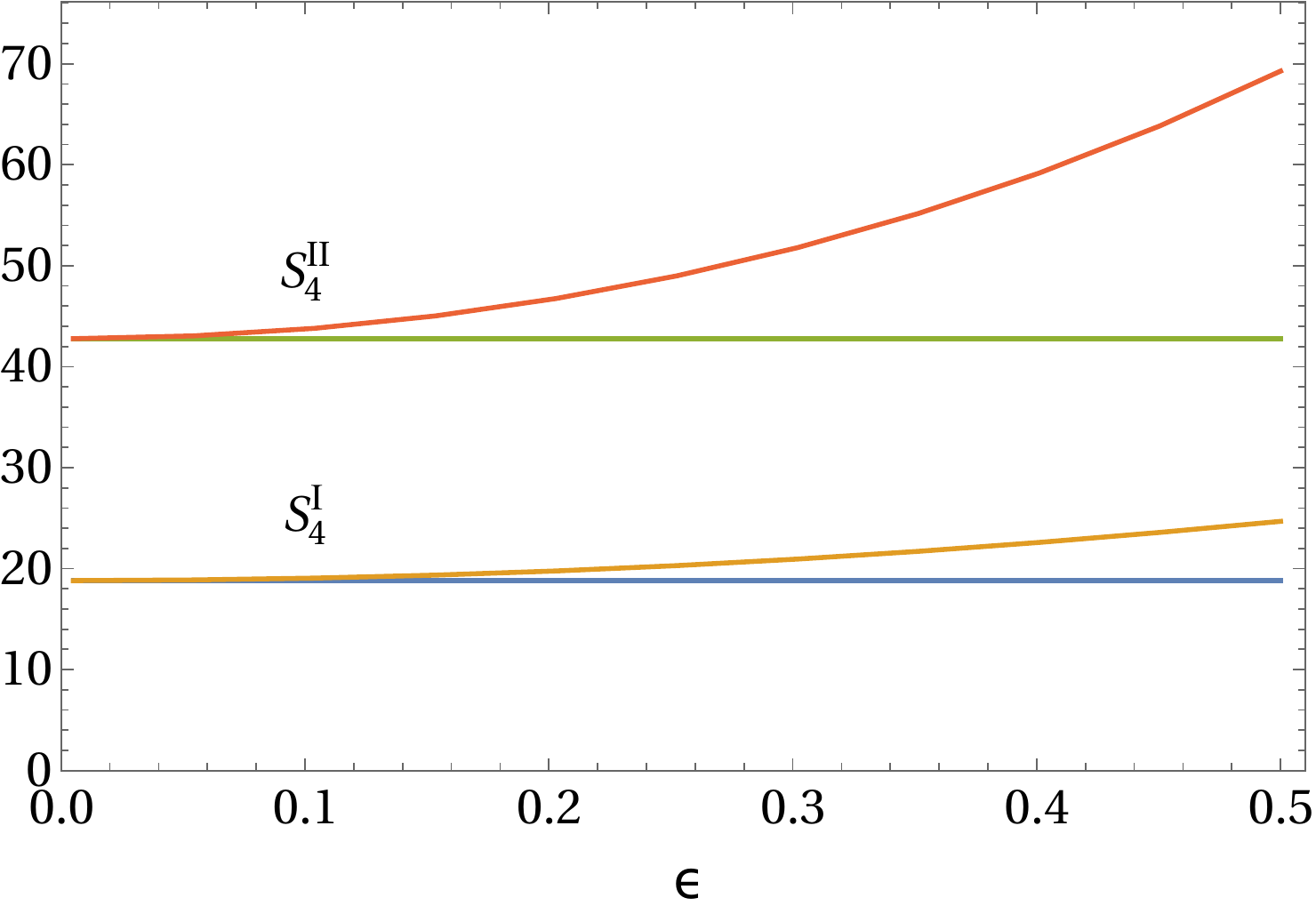}
\caption{\label{fig:03} Same as the Fig.~\ref{fig:02}, but the
  Gaussian shape of the power spectrum is assumed.
}
\end{figure}

In Fig.~\ref{fig:03}, the same comparisons as Fig.~\ref{fig:02} are made, but for the Gaussian shape of the spectrum. 
The narrow limits of $S_3^\mathrm{I}$, $S_4^\mathrm{I}$ and
$S_4^\mathrm{II}$ are the same as those in Fig.~\ref{fig:02} because they do not depend on the precise shape of the narrow spectrum. 
However, the narrow limits of the elements $S_3^\mathrm{II}$ and $S_3^\mathrm{III}$, as well as the exact numerical results of all the elements do depend on the precise shape of the spectrum.
As seen from the figures, the narrow limit approximation is slightly worse than that in the rectangular case for the same value of $\epsilon$.
The gradient of $S_3^\mathrm{III}$ as a function of $\epsilon$ is larger in the Gaussian case than in the rectangular case.
This results in the fact that $S_3^\mathrm{III}$ is non-negligible in comparison with $S_3^\mathrm{I}$ already for mildly small values of $\epsilon$.


\begin{figure}[tbp]
\centering
\includegraphics[width=.48\textwidth]{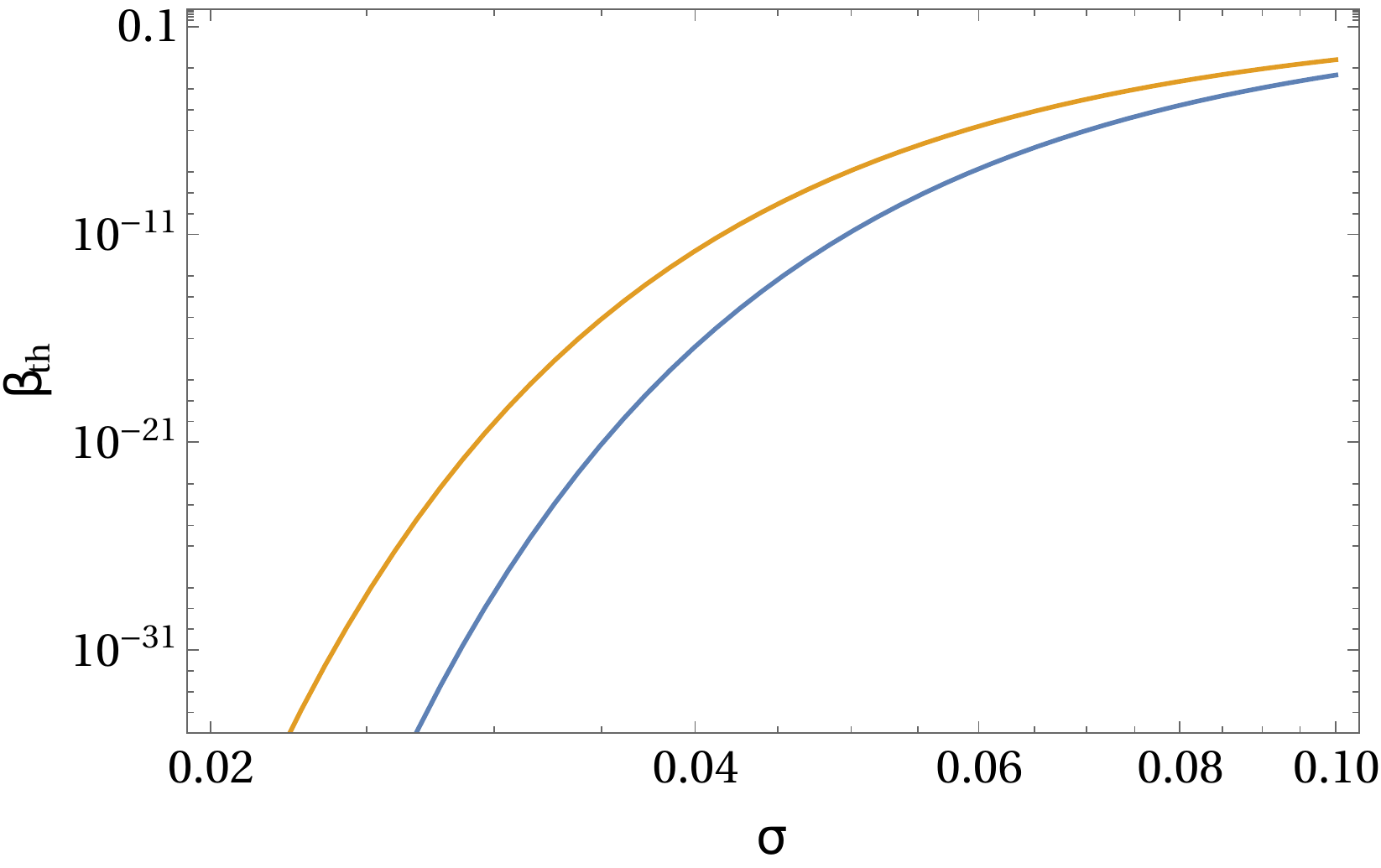}
\hfill
\includegraphics[width=.48\textwidth]{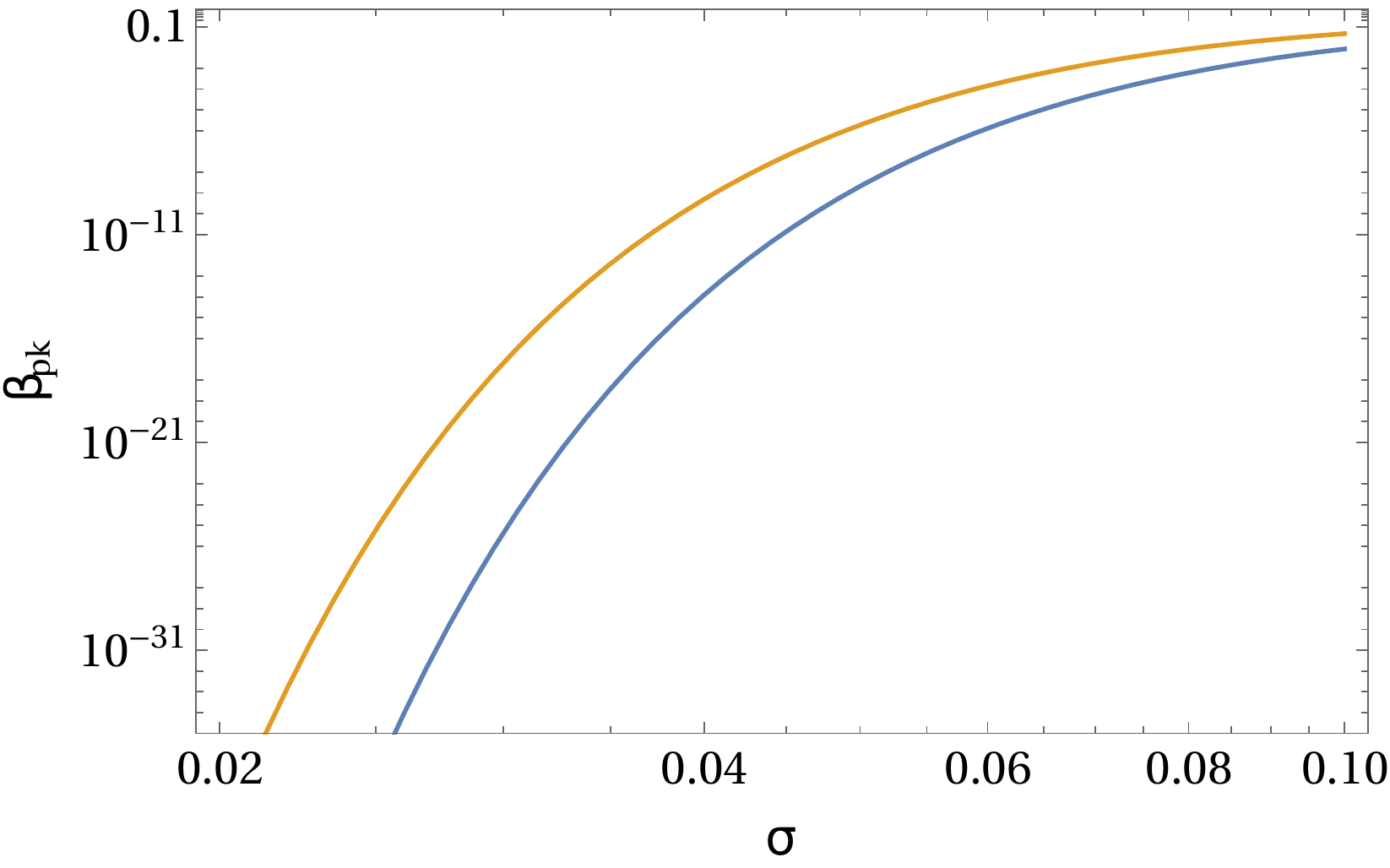}
\caption{\label{fig:04} Left panel: The mass fraction as a function of
  amplitude $\sigma$ of the primordial spectrum with the threshold
  model. The narrow limit of the spectral shape is assumed. The lower
  (blue) curve corresponds to the Gaussian initial condition and the
  upper (orange) curve corresponds to the non-Gaussian initial
  condition given by a model of Eq.~(\ref{eq:6-12}) with the narrow
  limit of spectrum and parameters are given by
  $\varDelta_\mathrm{c}=1/3$,
  $f_\mathrm{NL}^\mathrm{eff} = g_\mathrm{NL} = \tau_\mathrm{NL} =
  0.2$, and $k_0R = \sqrt{2}$. Right panel: Same as the right panel
  but with the peaks model.}
\end{figure}

In Fig.~\ref{fig:04}, we show the PBH mass fraction $\beta$ for the threshold model (left panel) and for the peaks model (right panel) in the narrow spectral shape limit, given by Eq.~(\ref{eq:6-12}). In this limit, the mass fraction does not depend on the precise shape of the spectrum, nor the precise value of the narrowness parameter. The adopted values of the parameters in the plot are 
$\varDelta_\mathrm{c}=1/3$,
$f_\mathrm{NL}^\mathrm{eff} = g_\mathrm{NL} = \tau_\mathrm{NL} = 0.2$,
and $k_0R = \sqrt{2}$. As can be seen in the figure, the PBH mass fraction is significantly enhanced for positive values of the non-Gaussian parameters $f_\mathrm{NL}^\mathrm{eff}$, $g_\mathrm{NL}$, and $\tau_\mathrm{NL}$.

The peaks model is known to predict a larger mass fraction than the threshold model for Gaussian initial conditions \cite{Yoo:2018kvb}. 
In the narrow spectral shape limit, however, the non-Gaussian factor in the exponent of Eq.~(\ref{eq:6-12}) does not depend on whether the formation criterion is determined by the threshold or peaks model, because it contains only $S_3^\mathrm{I}$ and $S_4^\mathrm{I}$, given by Eqs.~(\ref{eq:6-02a}) and (\ref{eq:6-11a}), respectively, which are independent of formation criteria. Therefore, the enhancement factor for the non-Gaussian case in comparison with the Gaussian case is the same for both panels.

\section{Conclusions}

In this paper, we considered the effect of primordial non-Gaussianities on the abundance of PBHs in the conventional PBH formation scenario in which a PBH is formed from a rare, large positive curvature perturbation at a radiation-dominated stage in the early universe. 
We first presented a general series form of the one-point distribution function with arbitrary non-Gaussianities. Then we derived an asymptotic formula for the PBH mass fraction at the time of formation in the high-peaks limit. The asymptotic formula for the threshold model for the PBH formation has been known. In this paper, we specifically derived the asymptotic formula for the peaks model and found that the enhancement factor is identical in both models if the non-Gaussianities are the same.

Next, we focused on the effect of skewness and kurtosis on the
abundance for a few specific models of non-Gaussianity. The skewness
and kurtosis are calculated from the bispectrum and trispectrum. For
the bispectrum, we considered non-Gaussian models which are described
by superpositions of specific forms of the bispectrum given by
Eq.~(\ref{eq:5-15}), which include popular models such as the
local-type, equilateral-type, folded-type and orthogonal-type
non-Gaussianities. The integral formulas to calculate the skewness
parameter in this series of models were derived in
Eq.~(\ref{eq:5-20a}) -- (\ref{eq:5-21}). For the trispectrum, we
considered a local-type model given by Eq.~(\ref{eq:5-30}). The
integral formulas of kurtosis parameter were derived in
Eqs.~(\ref{eq:5-34a}) and (\ref{eq:5-34b}).

The integral formulas to obtain skewness and kurtosis parameters are
numerically not so difficult to evaluate. However, to obtain
a clear, intuitive understanding of the results, we considered the
case when the primordial curvature perturbation power spectrum is
sharply peaked at a scale $k_0$ with the width $\Delta k=\epsilon k_0$
where $\epsilon$ ($<1$) is the narrowness parameter. As a result, the
precise shape of the spectrum gives only subdominant contributions to
the resulting skewness and kurtosis parameters, as shown in
Eqs.~(\ref{eq:6-02a}), (\ref{eq:6-04}), (\ref{eq:6-11a}) and
(\ref{eq:6-11b}). In the limit $\epsilon\to0$, we obtained an
asymptotic formula for the PBH mass fraction, Eq.~(\ref{eq:6-12}) for
various types of non-Gaussianities of the bispectrum and trispectrum.
In particular, we found that the non-Gaussianity parameters for
various types of the bispectrum are linearly combined to give an
effective $f_{\rm NL}^{\rm eff}$ that determines the PBH mass
fraction.

In this paper, we simply assume a linear relation between the
curvature perturbation and the density contrast in
Eq.~(\ref{eq:2-02}), and also assume formation models of PBHs with
threshold and peaks criteria with a fixed value of the threshold.
These assumptions are employed in order to make it possible for us to
analytically treat the problem. It is not obvious that these
simplistic assumptions hold in realistic situations
\cite{Riccardi:2021rlf,Young:2022phe,Musco:2008hv}. However, since our
non-Gaussian corrections in the high-peaks limit change the abundance
only through a multiplicative factor, as given by Eq.~(\ref{eq:4-06}),
whose form is common to the two different formation models, namely,
the threshold model and the peaks model. Therefore, one may expect
that this non-Gaussian factor in the high-peaks limit is universal,
not depending on the details of the formation models that can be more
complicated than those we assume in this paper. Unfortunately,
however, it is difficult to prove this expectation within the context
of this paper, as the derivation of our formula in the case of the
peaks model, for instance, is already complicated enough. Therefore,
we leave this issue for future studies.
  
We systematically studied the effect of non-Gaussianities on the mass
fraction of PBHs at the time of formation at the level of bispectrum
and trispectrum. We only considered the one-point distribution
function of the curvature perturbation, we have no clue about
non-Gaussianity effects on the spatial distribution of PBHs. The
effects of bispectrum and trispectrum on the PBH clustering are
investigated and discussed in
Refs.~\cite{Tada:2015noa,Suyama:2019cst,Matsubara:2019qzv}. It seems
there has been no further systematic analysis of non-Gaussianity
effects on the spatial distribution of PBHs in general situations.
This issue certainly deserves further studies.

\acknowledgments

We thank Chris Byrnes for useful conversations at the early stage of
this work. This work was supported by JSPS KAKENHI Grants
Nos.~JP19K03835 (TM), 21H03403 (TM), 19H01895 (MS), 20H04727(MS), and
20H05853(MS).
\appendix
\section{High-peaks limit of the peaks model with non-Gaussianity
  \label{app:A01}
}

In this Appendix, we derive the formula of Eq.~(\ref{eq:4-05}) for the
mass fraction in the high-peaks limit of the peaks model with
primordial non-Gaussianity. We rely on the fact that the number of
peaks above a threshold $\nu$ asymptotically approach to the Euler
characteristic of the three-dimensional body of the regions where
$\varDelta_R > \nu\sigma$ is satisfied. The expected value of Euler
characteristic in non-Gaussian fields above a threshold $\nu$ is
formally given by
\cite{Matsubara:1994wn,Matsubara:1995wj,Matsubara:2020fet}
\begin{equation}
  n_\chi(\nu) =
  \left\langle
    \exp\left(
      \sum_{n=3}^\infty \frac{1}{n!} \sum_{\mu_1,\cdots,\mu_n}
      M^{(n)}_{\mu_1\cdots\mu_n}
      \frac{\partial^n}{\partial A_{\mu_1}\cdots\partial A_{\mu_n}}
    \right)
    F_\chi(\bm{A},\nu)
  \right\rangle_\mathrm{G}.
  \label{eq:a-01}
\end{equation}
Various quantities in this equation are defined below in order. The
10-dimensional vector of variables $\bm{A}$ is composed of normalized
variables,
\begin{equation}
  \bm{A} =
  \left(
    \alpha, y_i, Z_{ij}
  \right),
  \label{eq:a-02}
\end{equation}
with $1\leq i \leq j \leq 3$, where
\begin{equation}
  \alpha = \frac{\varDelta_R}{\sigma_0},\quad
  y_i = \sqrt{3}\,\frac{\partial_i\varDelta_R}{\sigma_1},\quad
  Z_{ij} =
  \frac{3}{\gamma}
  \frac{\partial_i\partial_j\varDelta_R}{\sigma_2}
  + \delta_{ij} \frac{\varDelta_R}{\sigma_0},
  \label{eq:a-02-1}
\end{equation}
$\partial_i=\partial/\partial x_i$ is a coordinate derivative, and
\begin{equation}
  \gamma = \frac{{\sigma_1}^2}{\sigma_0\sigma_2}
  \label{eq:a-03}
 \end{equation}
is a spectral parameter, and $\sigma_j$ is defined by
Eq.~(\ref{eq:3-05}). The $n$-point cumulants of the variables $\bm{A}$
are denoted by 
\begin{equation}
  M^{(n)}_{\mu_1\cdots\mu_n} =
  \left\langle A_{\mu_1}\cdots A_{\mu_n} \right\rangle_\mathrm{c}.
  \label{eq:a-04}
\end{equation}
A local function for the number density of the Euler characteristic is
given by
\begin{equation}
  F_\chi(\bm{A},\nu) =
  \left(\frac{\sigma_1}{\sqrt{3}\sigma_0}\right)^3
  \Theta(\alpha-\nu)
  \delta_\mathrm{D}^3(\bm{y})
  \det(\alpha I - Z),
  \label{eq:a-05}
\end{equation}
where $\Theta(x)$ is a step function and $\delta_\mathrm{D}^3(\bm{y})$
is the 3-dimensional Dirac's delta function, and $I$ is the
$3\times 3$ unit matrix. Finally, we define a Gaussian average by
\begin{equation}
  \langle\cdots\rangle_\mathrm{G} =
  \frac{1}{(2\pi)^5\sqrt{\det M}}
  \int d^{10}A
  \cdots
  \exp\left(-\frac{1}{2}\bm{A}^\mathrm{T}{M}^{-1}\bm{A}\right),
  \label{eq:a-06}
\end{equation}
where $M = (M^{(2)}_{\mu_1\mu_2})$ is a $10\times 10$ matrix of
two-point cumulants. The two-point cumulants are explicitly given by
\begin{align}
  &
    \left\langle \alpha^2 \right\rangle = 1, \quad
    \left\langle \alpha \eta_i \right\rangle = 0, \quad
    \left\langle \alpha Z_{ij} \right\rangle = 0, \quad
    \left\langle y_i y_j \right\rangle = \delta_{ij}, \quad
    \left\langle \eta_i Z_{jk} \right\rangle = 0,
  \nonumber\\
  &
    \left\langle Z_{ij} Z_{kl} \right\rangle =
    \frac{3}{5\gamma^2}
    \left(
    \delta_{ij}\delta_{kl} + \delta_{ik}\delta_{jl} + \delta_{il}\delta_{jk}
    \right)
    - \delta_{ij}\delta_{kl}.
  \label{eq:a-07}
\end{align}

Expanding the operators in the exponent of Eq.~(\ref{eq:a-01}), we
have the following type of factor in the expansion:
\begin{equation}
  \left\langle
    \frac{\partial^{m_0}}{\partial \alpha^{m_0}}
    \frac{\partial^{m_1}}
    {\partial y_{i_1'}\cdots\partial y_{i_{m_1}'}}
    \frac{\partial^{m_2}}
    {\partial Z_{i_1j_1}\cdots\partial Z_{i_{m_2}j_{m_2}}}
    F_\chi
  \right\rangle_\mathrm{G}.
  \label{eq:a-08}
\end{equation}
This factor is equivalent to \cite{Matsubara:2020fet}
\begin{equation}
  \left(\frac{\sigma_1}{\sqrt{3}\sigma_0}\right)^3
  \int_\nu^\infty\!d\nu\,
  \left(-\frac{d}{d\nu}\right)^{m_0}
  \left[
    \left\langle
      \delta_\mathrm{D}(\alpha-\nu)
    \right\rangle_\mathrm{G}
    \left\langle
      \frac{\partial^{m_1} \delta_\mathrm{D}^3(\bm{y})}
      {\partial y_{i_1'}\cdots\partial y_{i_{m_1}'}}
    \right\rangle_\mathrm{G}
    \left\langle
      \frac{\partial^{m_2} \det(\nu I - Z)}
      {\partial Z_{i_1j_1}\cdots\partial Z_{i_{m_2}j_{m_2}}}
    \right\rangle_\mathrm{G}
  \right],
  \label{eq:a-09}
\end{equation}
because the three kinds of variables $\alpha$, $y_i$ and $Z_{ij}$ are
mutually independent in the Gaussian averages with two-point
cumulants. The Gaussian distribution functions of the variables
$\alpha$ and $y_i$ are given by
\begin{equation}
  P^{(0)}_\mathrm{G}(\alpha) =
  \frac{e^{-\alpha^2/2}}{\sqrt{2\pi}}, \quad
  P^{(1)}_\mathrm{G}(\bm{y}) =
  \frac{e^{-|\bm{y}|^2/2}}{(2\pi)^{3/2}},
  \label{eq:a-10}
\end{equation}
and the first two factors in the square bracket reduce to
\begin{equation}
    \left\langle
      \delta_\mathrm{D}(\alpha-\nu)
    \right\rangle_\mathrm{G} =
    \frac{e^{-\nu^2/2}}{\sqrt{2\pi}},\quad
    \left\langle
      \frac{\partial^{m_1} \delta_\mathrm{D}^3(\bm{y})}
      {\partial y_{i_1'}\cdots\partial y_{i_{m_1}'}}
    \right\rangle_\mathrm{G} =
    \begin{cases}
      \displaystyle
      \frac{(-1)^{m_1/2} (m_1-1)!!}{(2\pi)^{3/2}}
      \delta_{(i_1'i_2'}\cdots\delta_{i_{m_1-1}'i_{m_1}')},
      & (m_1: \mathrm{even}), \\
      0,
      & (m_1: \mathrm{odd}),
      \end{cases}
  \label{eq:a-11}
\end{equation}
where parentheses in the subscript of Kronecker's delta indicate the
symmetrization of the indices inside them. For the last factor, we
have an identity for the $3\times 3$ matrices,
\begin{equation}
  \det(\nu I - Z) =\nu^3 - \nu^2\,\mathrm{tr}\,Z
  + \frac{\nu}{2}
  \left[
    \left(\mathrm{tr}\,Z\right)^2 - \mathrm{tr}\,\left(Z^2\right)
  \right] - \det Z.
  \label{eq:a-12}
\end{equation}
For a fixed number of derivatives in Eq.~(\ref{eq:a-08}),
$m\equiv m_0+m_1+m_2$, one can see that the dominant contribution in
the high-peaks limit, $\nu\rightarrow\infty$, is given by a case of
$m_0=m$, $m_1=m_2=0$. In this case, we have
$\langle\det(\nu I-Z)\rangle_\mathrm{G} = H_3(\nu)$ due to
Eqs.~(\ref{eq:a-07}) and (\ref{eq:a-12}), and Eq.~(\ref{eq:a-09})
reduces to
\begin{equation}
  \frac{1}{(2\pi)^2}
  \left(\frac{\sigma_1}{\sqrt{3}\sigma_0}\right)^3
  e^{-\nu^2/2} H_{m+2}(\nu)
  \label{eq:a-13}
\end{equation}
in the high-peaks limit. When the replacement
$H_{m+2}(\nu) \rightarrow H_2(\nu) \nu^m$ in the same limit is applied
for each term of the expanded series of Eq.~(\ref{eq:a-01}), the
resulting infinite series is resummed again. Thus we have
\begin{equation}
  n_\chi(\nu) \approx
  \frac{1}{(2\pi)^2}
  \left(\frac{\sigma_1}{\sqrt{3}\sigma_0}\right)^3
  e^{-\nu^2/2} H_2(\nu)
    \exp\left(
      \sum_{n=3}^\infty \frac{\nu^n}{n!}
      \left\langle\alpha^n\right\rangle_\mathrm{c}
    \right).
  \label{eq:a-14}
\end{equation}
Identifying the number density of peaks with the Euler number density
in the high-peaks limit, the mass fraction of peaks is given by
\begin{equation}
  \beta_\mathrm{pk} \approx
  (2\pi)^{3/2}R^3 n_\chi(\nu)
  \approx
  \frac{1}{\sqrt{2\pi}}
  \left(\frac{R\sigma_1}{\sqrt{3}\sigma}\right)^3
  (\nu^2-1) e^{-\nu^2/2}
  \exp\left(
    \sum_{n=3}^\infty \frac{\nu^n}{n!}
    \left\langle\alpha^n\right\rangle_\mathrm{c}
  \right).
  \label{eq:a-15}
\end{equation}
The cumulants of $\alpha$ are given by
$\langle \alpha^n \rangle_\mathrm{c} = \sigma^{n-2}S_n$ where $S_n$ is
the reduced cumulants defined by Eq.~(\ref{eq:4-01}), and the above
equation is equivalent to Eq.~(\ref{eq:4-06}).


\end{document}